\documentclass[aps,10pt,prd,groupedaddress,nofootinbib,notitlepage,eqsecnum,preprintnumbers]{revtex4-1}

\usepackage[utf8]{inputenc}
\usepackage{hyperref}
\usepackage{xcolor}
\usepackage{graphicx}
\usepackage{amsmath,amssymb,amsbsy,amstext,amsthm}
\usepackage{mathtools}
\usepackage{amsfonts}
\usepackage{comment}
\usepackage{bm}

\begin{document}
\title{Induced gravitational waves in a general cosmological background}

\author{\textsc{Guillem Dom\`enech$^{a}$}}
		\email{{domenech}@{thphys.uni-heidelberg.de}} 
\affiliation{$^{a}$\small{Institut f\"ur Theoretische Physik, Ruprecht-Karls-Universit\"at Heidelberg,\\ Philosophenweg 16, 69120 Heidelberg, Germany}}

\begin{abstract}
Gravitational waves are inevitably produced by second order terms in cosmological perturbation theory. Most notably, the so-called induced gravitational waves are a window to the small scales part of the primordial spectrum of fluctuations and a key counterpart to the primordial black hole scenario. However, semi-analytical solutions are only known for matter and radiation domination eras. In this paper, we present new analytic integral formulas for the induced gravitational waves on subhorizon scales in a general cosmological background with a constant equation of state. We also discuss applications to a peaked primordial scalar power spectrum and the primordial black hole scenario. 
\end{abstract}
 \maketitle

\section{Introduction\label{sec:intro}}

The forthcoming space-based gravitational wave (GW) detectors, such as LISA \cite{Audley:2017drz}, DECIGO \cite{Seto:2001qf,Yagi:2011wg}, AION/MAGIS \cite{Badurina:2019hst}, ET \cite{ET} and PTA \cite{Lentati:2015qwp,Shannon:2015ect,Arzoumanian:2015liz,Qin:2018yhy}, will open the field of gravitational wave cosmology. They will provide a unique opportunity to explore unobserved periods and test various models of the early universe (for a review see \textit{e.g.} \cite{Maggiore:1999vm,Sathyaprakash:2009xs,Guzzetti:2016mkm,Caprini:2018mtu} and references therein). Most of the information relevant for cosmology is likely to be encoded in a stochastic gravitational wave background (SGWB) due to the large number of unresolved sources \cite{Guzzetti:2016mkm,Christensen:2018iqi,Caprini:2018mtu}. Importantly, the cosmological SGWB differs from the astrophysical one in its strength, frequency range, spectral shape and statistical properties. Thus, it is in principle possible to extract any signal of cosmological origin once the astrophysical sources have been accurately modeled \cite{Regimbau:2011rp}.

The study of the cosmic microwave background (CMB) has already provided invaluable information on the initial conditions of the universe. In the scales probed by the CMB, the primordial universe was homogeneous and isotropic, filled with tiny adiabatic gaussian fluctuations with an almost scale invariant spectrum \cite{Akrami:2018odb,Aghanim:2018eyx}. These results strongly support inflation \cite{Brout:1977ix,Starobinsky:1979ty,Guth:1980zm,Sato:1980yn} as the paradigm for the very early universe. However, there is little known about the small scale region of the primordial spectrum, which does not affect the CMB, and that depends on the dynamics around the end of inflation. The SGWB might prove to be key to test those scales \cite{Assadullahi:2009jc,Bugaev:2010bb,Inomata:2018epa,Sato-Polito:2019hws}.

One remarkable fact is that there are very strong arguments in favor of the existence of a cosmological SGWB. Since gravity is a non-linear theory, these initial fluctuations set by inflation actually become a source of GWs at second order in cosmological perturbation theory \cite{Kodama:1985bj,Mukhanov:1990me,Noh:2004bc,Hwang:2007ni,Ananda:2006af,Baumann:2007zm}. This type of SGWB is referred to as the induced SGWB and it is has attracted much attention recently \cite{Alabidi:2012ex,Alabidi:2013wtp,Hwang:2017oxa,Kohri:2018awv,Cai:2018dig,Bartolo:2018rku,Inomata:2018epa,Yuan:2019udt,Inomata:2019zqy,Inomata:2019ivs,Chen:2019xse,Yuan:2019wwo,DeLuca:2019ufz,Tomikawa:2019tvi,Gong:2019mui,Inomata:2019yww,Yuan:2019fwv} since, in some cases, it can be larger than the SGWB generated from quantum fluctuations during inflation. 

From the observational side, the induced SGWB is an important probe of the thermal history of the universe \cite{Watanabe:2006qe} (also see \cite{Cui:2017ufi,Bettoni:2018pbl,Cui:2018rwi} and references therein for a similar discussion with cosmic strings) and the spectrum of primordial fluctuations \cite{Assadullahi:2009jc,Bugaev:2010bb,Inomata:2018epa}. Furthermore, the detailed shape of the induced SGWB spectrum highly depends on the non-gaussian nature of the primordial fluctuations \cite{Cai:2018dig}. In other words, the induced SGWB is sensitive to the three point function of primordial fluctuations. From the pure theoretical side, there is currently a debate on the reported gauge dependence on the generation of GWs at second order \cite{Hwang:2017oxa,Domenech:2017ems,DeLuca:2019ufz,Gong:2019mui,Inomata:2019yww,Yuan:2019fwv}, which directly affects the induced GW spectrum. We will not attempt to solve this issue here but we believe that, nevertheless, results of the paper will be useful as they can be translated to any desired gauge \cite{Hwang:2017oxa}.

The induced SGWB plays also a key role in the primordial black hole (PBH) scenario \cite{Sasaki:2016jop,Bird:2016dcv,Carr:2016drx,Garcia-Bellido:2017fdg}, which is a plausible candidate to dark matter without a particle physics nature (for a review see \textit{e.g.} \cite{Sasaki:2018dmp} and references therein). The main reason is that the PBH scenario requires that inflation generated huge fluctuations on scales not probed by the CMB, which later collapsed to black holes \cite{Sasaki:2018dmp}. Interestingly, the PBH scenario leaves traces in distortions of the CMB \cite{Nakama:2017xvq}, lensing \cite{Niikura:2017zjd,Zumalacarregui:2017qqd} and direct detection of GWs from PBHs mergers \cite{Sasaki:2016jop,Bird:2016dcv,Gow:2019pok}. In addition to that, it seems that if PBHs constitute a considerable fraction, if not all, of the dark matter, there must be a detectable induced SGWB counterpart \cite{Bugaev:2009zh,Saito:2009jt,Alabidi:2013wtp,Kohri:2018awv,Cai:2018dig,Bartolo:2018rku,Inomata:2018epa,Inomata:2019zqy,Inomata:2019ivs,Yuan:2019udt}.

Now, most of the analytical studies of the induced SGWB are restricted to either radiation or matter domination, otherwise numerical methods were needed \cite{Tomikawa:2019tvi}. While there are strong motivations to study the generation of the induced SGWB during radiation domination, matter domination has been considered first due to an enhancement of the amplitude of the SGWB and the unknown physics right after inflation \cite{Baumann:2007zm,Assadullahi:2009nf,Alabidi:2013wtp}. For example, if after the end of inflation, the inflaton oscillates around the minimum of its potential, the universe will undergo a phase of matter domination before the universe is (re)heated (assuming that the potential is approximately quadratic at the bottom). However, this is not necessarily the case. Quintessential inflation scenarios \cite{Wetterich:1987fm,Spokoiny:1993kt,Wetterich:1994bg,Peebles:1998qn,Brax:2005uf,Wetterich:2014gaa,Hossain:2014zma,Hossain:2014xha,Karananas:2016kyt,Agarwal:2017wxo,Geng:2017mic,Dimopoulos:2017zvq,Rubio:2017gty,Haro:2018zdb} (see also \cite{GarciaBellido:2011de,Casas:2017wjh} in the context of Higgs inflation) assume that, after inflation, the inflaton has a run away potential and the universe becomes kinetic dominated until (re)heating is achieved. Moreover, the equation of state of the universe is not necessarily constant throughout the whole evolution \cite{Byrnes:2018clq,Hajkarim:2019nbx} and neither it will be exactly $w=1/3$. The induced SGWB spectrum has the potential to distinguish these scenarios as a probe of the thermal evolution of the universe. Here we will study the generation of induced GWs in a general background with a constant equation of state and their applications to the PBH scenario.

This paper is organized as follows. In Sec.~\ref{sec:cosmobackground} we review the cosmology in a FRLW background with a constant equation of state and the calculation of induced GWs in such general background. In Sec.~\ref{sec:analyticalcalculation} we present new analytical formulas for the induced gravitational waves kernel on subhorizon scales. In Sec.~\ref{sec:PBH} we discuss its application to a peaked primordial spectrum in cosmological backgrounds such as kinetic domination (or stiff equation of state). Sec.~\ref{sec:conclusions} is dedicated to the conclusions of this work. Details of the calculations and properties of the special functions are provided in the appendices. In particular, Apps.~\ref{App:bessel}, \ref{App:integralbessel} and \ref{App:legendre} are extensively dedicated to the mathematical properties of Bessel and associated Legendre functions.

\section{Scalar induced gravitational waves review \label{sec:cosmobackground}}

Gravitational waves are unavoidably generated in general by second order terms in cosmological perturbation theory. To see that, we consider a perturbed flat FLRW metric in the zero shear (or Poisson) gauge,
\begin{align}\label{eq:poisson}
ds^2&=a^2(\tau)\left[-(1+2\Phi)d\tau^2+\left(\delta_{ij}-2\Psi\delta_{ij}+2h_{ij}\right)dx^idx^j\right]\,,
\end{align}
where $a(\tau)$ is the scale factor, $\tau$ is conformal time, $\Phi$ and $\Psi$ are the gravitational scalar potentials and $h_{ij}$ are the (transverse-traceless) tensor perturbations or gravitational waves.\footnote{At linear order the identification of tensor modes with GWs is straightforward as they are gauge invariant. However, at second order is not so clear anymore. For more on this see \cite{Arroja:2009sh,Hwang:2017oxa}.} As usual, we have neglected vector perturbations since they decay. We also assume that the dominant matter content of the universe is given by a perfect fluid, \textit{i.e.} its energy momentum tensor reads
\begin{align}
T_{\mu\nu}=(\rho+p)u_{\mu}u_{\nu}+pg_{\mu\nu}\,,
\end{align}
where $\rho$ and $p$ are respectively the energy density and pressure of the fluid, $u_{\mu}$ is its 4-velocity and $g_{\mu\nu}$ is the spacetime metric \eqref{eq:poisson}. For a constant equation of state,
\begin{align}
w=p/\rho={\rm constant}\,,
\end{align}
the Friedmann equations yield
\begin{align}\label{eq:background}
a(\tau)=a(\tau_i)\left(\frac{\tau}{\tau_i}\right)^{\frac{2}{\left(1+3w\right)}}\quad,\quad{\cal H}&\equiv\frac{a'}{a}=\frac{2}{\left(1+3w\right)\tau}\,,
\end{align}
where $\tau_i$ is a reference time and a prime denotes derivative with respect to conformal time. Recall that $w=0,1/3$ respectively correspond to matter and radiation domination. In this paper we will focus on $0<w\leq1$. That is all we need to know of the cosmological background. More details can be found in App.~\ref{App:first order}.

At linear order in cosmological perturbation theory, the gravitational potential and the tensor modes in absence of anisotropies\footnote{For example, due to massive neutrinos, GWs will experience a further damping on very small scales \cite{Mangilli:2008bw}.} respectively satisfy the following equations of motion \cite{Kodama:1985bj,Mukhanov:1990me},
\begin{align}
\Phi''+3(1+w){\cal H}\Phi'-w\Delta\Phi=0\,
\quad {\rm and} \quad
h_{ij}''+2{\cal H}h_{ij}'-\Delta h_{ij}=0\,,
\end{align}
where $\Delta\equiv\delta^{ij}\partial_i\partial_j$. Working in Fourier space (see App.~\ref{app:fourier} for the conventions used) and using Eq.~\eqref{eq:background} we arrive at
\begin{align}
\Phi''+6\frac{1+w}{1+3w}\frac{1}{\tau}\Phi'+wk^2\Phi=0\quad {\rm and}\quad h_\lambda''+\frac{4}{1+3w}\frac{1}{\tau}h_\lambda'+k^2h_\lambda=0\,,
\end{align}
where $k$ is the wavenumber and $\lambda$ is the GWs polarization, \textit{e.g.} $\lambda=\{R, L\}$, where $R$ and $L$ respectively refer to right and left polarizations. General solutions to these equations are given by \cite{Baumann:2007zm}
\begin{align}\label{eq:alpha}
\Phi(x)=(\sqrt{w}x)^{-\alpha}\left[C_1(k)J_\alpha(\sqrt{w}x)+C_2(k)Y_\alpha(\sqrt{w}x)\right]\quad,\quad \alpha=\frac{5+3w}{2(1+3w)}
\end{align}
and
\begin{align}\label{eq:beta}
h_\lambda(x)=x^{-\beta}\left[\tilde C_1(k)J_\beta(x)+\tilde C_2(k)Y_\beta(x)\right]\quad,\quad \beta=\frac{3}{2}\frac{1-w}{1+3w}\,,
\end{align}
where $x\equiv k\tau$ and $J_\nu$ and $Y_{\nu}$ respectively are Bessel functions of the first and second kind. Also, it will important later to realize that $\alpha=1+\beta$. These solutions simply mean that after the modes enter the horizon they will start damped oscillations due to pressure. We will be interested in the range
\begin{align}
1\geq w > 0 \quad \Rightarrow \quad 1\leq \alpha< 5/2 \quad,\quad 0\leq \beta< 3/2\,,
\end{align}
for the following reasons. On one hand, we exclude the case $w=0$ since solutions to Eq.~\eqref{eq:alpha} are not given by Bessel functions but by a constant and a decaying mode, although one can formally take the limit $w\to0$. Furthermore, an early period of matter domination ($w=0$) cannot last longer than a certain amount of time, otherwise non-linearities become important \cite{Assadullahi:2009nf,Alabidi:2013wtp,Inomata:2019ivs,Inomata:2019zqy}. On the other hand, if $w<0$ then we get modified Bessel functions as solutions of Eq.~\eqref{eq:alpha} and the analytical calculation of Sec.~\ref{sec:analyticalcalculation} would be somewhat more complicated. We leave the extension to $w<0$ for future work.

The initial conditions set by inflation are constant on superhorizon scales ($x\ll1$). Thus, using the fact that $\lim_{x\to 0}x^{-\nu}{J_\nu(x)}={2^{-\nu}}/{\Gamma[1+\nu]}$
while $x^{-\nu}Y_\nu\propto x^{-2\nu}$ diverges for $x\to 0$, we extract
\begin{align}
C_1=2^\alpha\Gamma[1+\alpha]\Phi_{\rm prim}(k) \quad{,}\quad \tilde C_1=2^\beta\Gamma[1+\beta]h_{\lambda,{\rm prim}}(k)\,,
\end{align}
and $C_2=\tilde C_2=0$, where the subindex ``prim'' refers to its primordial value. We can now relate the value of the gravitational potential to that of the curvature perturbation on superhorizon scales, which is the quantity that remains constant, by \cite{Kodama:1985bj,Mukhanov:1990me,Kohri:2018awv}
\begin{align}
\Phi_{\rm prim}(k)=\frac{3+3w}{5+3w}{\cal R}_{{\rm prim}}(k)\,.
\end{align}

Turning now our attention to GWs generated in the early universe, they will lead to a stochastic background due to the large number of unresolved sources. The amplitude of such stochastic background is related to the energy density of GWs, which is defined\footnote{We find the value of the energy density of GWs by expanding Einstein equations to second order.} on sub-horizon scales by \cite{Maggiore:1900zz,Watanabe:2006qe}
\begin{align}
\rho_{\rm GWs}=\frac{M_{pl}^2}{4a^2}\langle h'_{ij}h'_{ij}\rangle\approx\frac{M_{pl}^2k^2}{4a^2}\overline{\langle h_{ij}h_{ij}\rangle}\,.\nonumber
\end{align}
In the last equality we used that deep inside the horizon GWs oscillate and, therefore, $h'_{ij}\sim k h_{ij}$. The large line on top of the expectation value refers to oscillation average\footnote{If the GWs have single frequency and oscillate rapidly we can take integral over half period and divide by $\pi$. This is not a good approximation on scales close to (or beyond) the horizon.} and estimates the observed amplitude of the stochastic background \cite{Maggiore:1900zz,Kohri:2018awv}.

It is convenient to define the relative spectral density of GWs in a given background per $\ln k$ as
\begin{align}\label{eq:spectraldensity}
\Omega_{\rm GWs}(k,\tau)=\frac{\rho_{\rm GWs}(k,\tau)}{3H^2M_{pl}^2}=\frac{k^2}{12a^2H^2}\overline{P_h(k,\tau)}
\quad{\rm where}\quad
\rho_{\rm GWs}(k,\tau)=\frac{d\rho_{\rm GWs}(\tau)}{d\ln k}\,,
\end{align}
and we have used that
\begin{align}
\langle h_{ij}(x) h_{ij}(x)\rangle=\int \frac{d^3k}{(2\pi)^3} \langle h_R(k,\tau)h_R(k,\tau)\rangle+\langle h_L(k,\tau)h_L(k,\tau)\rangle=2\int \frac{d^3k}{(2\pi)^3} \langle h_R(k,\tau)h_R(k,\tau)\rangle\equiv\int d\ln k \,P_h(k,\tau)\,,
\end{align}
where we chose the left (L) and right (R) polarizations for convenience as we assume no parity violation.

From the non-observation of tensor modes in the CMB we have that $r=P_h/P_{\cal R}<0.07$ \cite{Akrami:2018odb}. Thus, if we assume that the primordial tensor spectrum is almost scale invariant with a slight red tilt, then primordial GWs are practically unobservable by future GW detectors \cite{Guzzetti:2016mkm}. For this reason, we will not consider the evolution of primordial GWs and, thus, we will set $h_{\lambda,{\rm prim}}(k)=0$ without loss of generality. However, see \cite{Garcia-Bellido:2017aan,Unal:2018yaa} for the possible effects of non-zero primordial tensor modes. Then, in what follows we will be only concerned in the sourcing of GWs $h_{ij}$ by first order squared gravitational potential $\Phi$ terms. Furthermore, we will neglect the details in the transfer function of GWs due to changes in the relativistic degrees of freedom \cite{Watanabe:2006qe}. We will also assume that GWs propagate freely to us after they are formed, \textit{e.g.} there is no strong magnetic field modifying the propagation of GWs as in Ref.~\cite{Bamba:2018cup}.

\subsection{Induced tensor modes at second order\label{sec:secondorder}}

At second order in cosmological perturbation theory, scalar and tensor modes mix \cite{Noh:2004bc}. The solutions to the first order equations then source the second order ones. Looking at the transverse-traceless part of the $ij$ component of Einstein equations to second order we find that scalar squared terms source tensor modes by \cite{Baumann:2007zm,Domenech:2017ems}
\begin{align}\label{eq:hij}
h_{ij}''+2{\cal H}h_{ij}'-\Delta h_{ij}=\widehat{TT}^{ab}_{ij}\left[4\partial_a\Phi\partial_b\Phi+\frac{8}{3\left(1+w\right)}\partial_a\left(\Phi+\Phi'/{\cal H}\right)\partial_b\left(\Phi+\Phi'/{\cal H}\right)\right]\,,
\end{align}
where $\widehat{TT}^{ab}_{ij}$ is the transverse-traceless projector.\footnote{The transverse-traceless projector is explicitly given by
\begin{align}
\widehat{TT}_{ij}\,^{ab}=&\left(\delta_{i}^{(a}-\partial_i\partial^{(a}\Delta^{-1}\right)\left(\delta_j^{b)}-\partial_j\partial^{b)}\Delta^{-1}\right)-\frac{1}{2}\left(\delta_{ij}-\partial_i\partial_j\Delta^{-1}\right)\left(\delta^{ab}-\partial^a\partial^b\Delta^{-1}\right)\nonumber\,.
\end{align}
} We have neglected source terms containing scalar-tensor and tensor-tensor contributions as they are subdominant \cite{Gong:2019mui}.
 Now, in Fourier space Eq.~\eqref{eq:hij} reduces to
\begin{align}\label{eq:hij1}
h_{\lambda}''+2{\cal H}h_{\lambda}'+k^2h_{\lambda}=\,s_\lambda(\mathbf{k})\,,
\end{align}
where
\begin{align}
s_\lambda(\mathbf{k})=4\int \frac{d^3q}{(2\pi)^3}e_\lambda^{ij}(k)q_iq_j\left\{\Phi(\mathbf{q})\Phi(\mathbf{k}-\mathbf{q})+\frac{2}{3\left(1+w\right)}\left[\Phi(\mathbf{q})+\frac{\Phi'(\mathbf{q})}{\cal H}\right]\left[\Phi(\mathbf{k}-\mathbf{q})+\frac{\Phi'(\mathbf{k}-\mathbf{q})}{\cal H}\right]\right\}\,,
\end{align}
and $e_{\lambda}^{ij}$ is the polarization tensor of GWs (see App.~\ref{app:fourier} for details in the derivation). For instance, if we choose the left and right polarizations where $\lambda=\{R,L\}$, where
 $e_\lambda^{ij}(\mathbf{k})e^{\lambda'}_{ij}(-\mathbf{k})=\delta^{\lambda\lambda'}$ and $e_{R,L}^{ij}(\mathbf{k})k_j=0$, we have that
\begin{align}
e_{R,L}^{ij}(\mathbf{k})q_iq_j=\frac{1}{2}q^2\sin^2\theta\,,
\end{align}
where $\theta$ is the angle between $\mathbf{k}$ and $\mathbf{q}$. We can formally solve Eq.~\eqref{eq:hij1} by the Green's function method, which yields \cite{Baumann:2007zm}
\begin{align}
h_\lambda(k)=\int_{x_0}^x d\tilde x \,k^{-2}s_\lambda(\mathbf{k},\tilde x)G(x,\tilde x)
\quad{\rm where}
\quad
G(x,\tilde x)\equiv\frac{y_1(\tilde x)y_2(x)-y_2(\tilde x)y_1(\tilde x)}{y_1(\tilde x)\partial_{\tilde x}y_2(x)-y_2(\tilde x)\partial_{\tilde x}y_1(\tilde x)}\,,
\end{align}
$y_1$, $y_2$ are solutions to the homogeneous equation Eq.~\eqref{eq:beta}
and we assumed proper boundary conditions for the Green's function.

In the cases under study the power spectrum of induced GWs\footnote{We should actually use 
\begin{align}
h'_\lambda(k,\tau)=k\int d\tilde x \,k^{-2}s_\lambda(\mathbf{k},\tilde x)\partial_x G(x,\tilde x)\nonumber
\end{align}
for the calculation of the GWs energy density. However, $h'_\lambda(k)\approx k h_\lambda(k)$ on sub-horizon scales.} takes a particularly compact form \cite{Kohri:2018awv},
\begin{align}\label{eq:pgamma}
\overline{P_h(k,\tau)}=2\int_0^\infty dv \int_{|1-v|}^{1+v} du\left[\frac{4v^2-\left(1+v^2-u^2\right)^2}{4uv}\right]^2 P_{\cal R}(kv)P_{\cal R}(ku)\overline{I^2}(v,u,x)\,,
\end{align}
where $P_{\cal R}$ is the primordial curvature power spectrum, $I(u,v,x)$ is the kernel, that we will shortly define in detail, and we used
\begin{align}
v\equiv q/k\quad,\quad u\equiv|\mathbf{k}-\mathbf{q}|/k\,.
\end{align}
In this derivation we have assumed gaussian primordial fluctuations and we have used that
\begin{align}
\langle \Phi_{\rm prim}(k)\Phi_{\rm prim}(k')\rangle=(2\pi)^3\frac{2\pi^2}{k^3}P_{\Phi,{\rm prim}}(k)\delta(k+k')\,.
\end{align}
Including non-gaussianity should be straightforward but it would be out of the scope of this paper.

Coming back to the kernel in Eq.~\eqref{eq:pgamma}, it can be written after setting $x_0=0$ as
\begin{align}\label{eq:kernel}
I(v,u,x)\equiv 2\left(\frac{3+3w}{5+3w}\right)^2\int_{0}^x d\tilde x \,G(x,\tilde x) f(\tilde x, u, v)\,,
\end{align}
where the green function, after using Eq.~\eqref{eq:beta}, simplifies to
\begin{align}\label{eq:green2}
G(x,\tilde x)=\frac{\pi}{2}\frac{\tilde x^{1+\beta}}{x^{\beta}}\left[J_\beta(\tilde x)Y_\beta(x)-Y_\beta(\tilde x) J_\beta(x)\right]\,,
\end{align}
and the source term is given by
\begin{align}\label{eq:source}
f(\tilde x, u, v)=&\frac{4^\alpha}{6\alpha}\Gamma^2[1+\alpha]\frac{1+3w}{1+w}\left(uvw\tilde x^2\right)^{1-\alpha}\nonumber\\&
\times\left[J_{\alpha-1}(u\sqrt{w}\tilde x)J_{\alpha-1}(v\sqrt{w}\tilde x)+\frac{3(1+w)}{2}J_{\alpha+1}(u\sqrt{w}\tilde x)J_{\alpha+1}(v\sqrt{w}\tilde x)\right]\,,
\end{align}
where we used Eq.~\eqref{eq:alpha} and we replaced $J_\alpha$ in terms of $J_{\alpha\pm1}$ (see App.~\ref{App:bessel} for more details). 

So far, we have reviewed the analytical integral solution for the spectrum of induced GWs \cite{Baumann:2007zm,Kohri:2018awv}. However, for practical purposes this form of the solution does not yield much physical insight as there is no full analytical expression for such integrals. Thus, only particular cases where the kernel integral \eqref{eq:kernel} is doable, \textit{e.g.} matter or radiation domination, have been studied or numerical methods were used \cite{Tomikawa:2019tvi}. In the next section we present analytical results for the kernel in a general background with constant equation of state on subhorizon scales.

\section{Semianalytical calculation\label{sec:analyticalcalculation}}

In order to greatly simplify the calculation of the induced GWs, one would ideally like to have an analytical expression for the kernel, since it is the heaviest calculation to do numerically as it encompasses the product of three Bessel functions. Using the kernel Eq.~\eqref{eq:kernel} derived in Sec.~\ref{sec:secondorder} and plugging in Eqs.~\eqref{eq:green2} and \eqref{eq:source}, we can simplify the expression of the kernel to
\begin{align}\label{eq:kernel2}
I(x,u,v)=4^{\beta}\frac{3\pi}{2\alpha^3}\frac{1+w}{1+3w}\Gamma^2[\beta+2]&\left(uvwx\right)^{-\beta}\left\{Y_\beta(x){\cal I}^x_J(u,v,w)-J_\beta(x){\cal I}^x_Y(u,v,w)\right\}
\end{align}
where we used that $\alpha=1+\beta$ and we defined
\begin{align}
{\cal I}^x_{J,Y}
(u,v,w)\equiv\int_0^{x} d\tilde x \tilde x^{1-\beta}
\left\{	
\begin{aligned}
	J_\beta(\tilde x)\\
	Y_\beta(\tilde x)
\end{aligned}
\right\}
\left[J_{\beta}(u\sqrt{w}\tilde x)J_{\beta}(v\sqrt{w}\tilde x)+\frac{3(1+w)}{2}J_{\beta+2}(u\sqrt{w}\tilde x)J_{\beta+2}(v\sqrt{w}\tilde x)\right]\,.
\end{align}
In this way, we have been able to write the integrals in terms of Bessels functions of order $\beta$ and $\beta+2$ only.

We will not be able to carry out this integral for general values of $x$ but we can obtain analytical results
for the scales of interest which are subhorizon, \textit{i.e.} $x\gg1$. In this case, we can approximate the integral by sending the upper limit $x\to\infty$. This will capture the leading order behavior of the kernel since $J_\beta(x)$ and $Y_\beta(x)$ already decay as $1/\sqrt{x}$ when $x\gg 1$. Although corrections from a finite upper integration limit can be computed,\footnote{We can split the integral
\begin{align}
\int_0^x=\int_0^\infty-\int_x^\infty\,,\nonumber
\end{align}
and then Taylor expand the integrand inside the second integral for large $\tilde x$ to estimate the next to leading order correction.} they will be suppressed by a further $1/x$. Interestingly, this type of integrals appeared in high-energy physics phenomenology to estimate absorption effects. Conveniently, Gervois and Navelet in Ref.~\cite{threebesselI} presented an analytical formula for the integrals ${\cal I}^\infty_{J,Y}(u,v,w)$ in terms of associated Legendre polynomials and Legendre polynomials on the cut. They extended the original result of Bailey in Ref.~\cite{originalbessel} to the full parameter space, who related the integral of three Bessel functions with Appell's fourth function by writing the Bessel functions as power series and then integrating. We will apply their results to the induced GWs. Details of the original formulas can be found in App.~\ref{App:integralbessel}.

We find that the integrals in Eq.~\eqref{eq:kernel2} are given by
\begin{align}\label{eq:IJ}
{\cal I}^\infty_J(u,v,w)=&\frac{2^{-\beta}(wZ)^{\beta-1/2}}{\sqrt{\pi uvw}}\left[\mathsf{P}^{-\beta+1/2}_{\beta-1/2}(y)+\frac{3(1+w)}{2}\mathsf{P}^{-\beta+1/2}_{\beta+3/2}(y)\right]\Theta(u+v-w^{-1/2})
\end{align}
and
\begin{align}\label{eq:IY}
{\cal I}^\infty_Y(u,v,w)=-\frac{2^{-\beta+1}w^{\beta-1/2}}{\pi\sqrt{\pi uvw}}&\Bigg\{Z^{\beta-1/2}\left[
	\mathsf{Q}^{-\beta+1/2}_{\beta-1/2}(y)+\frac{3(1+w)}{2}\mathsf{Q}^{-\beta+1/2}_{\beta+3/2}(y)\right]\Theta(u+v-w^{-1/2})\\&
	-\tilde Z^{\beta-1/2}\left[{\cal Q}^{-\beta+1/2}_{\beta-1/2}(\tilde y)+3(1+w){\cal Q}^{-\beta+1/2}_{\beta+3/2}(\tilde y)\right]\Theta(w^{-1/2}-u-v)\Bigg\}\,.
\end{align}
Let us explain in a bit of detail the notation in Eqs.~\eqref{eq:IJ} and \eqref{eq:IY}. First, we note that the integral may become singular for the resonant case where $u+v=w^{-1/2}$. This is when the frequency of two scalar modes matches that of the tensor mode and a narrow resonance occurs. Thus, to distinguish between the cases $u+v>w^{-1/2}$ and $u+v<w^{-1/2}$ we have introduced the Heavyside theta function $\Theta(z)$ which is $1$ when $z>0$ and $0$ otherwise. Secondly, all functions have been defined so that they are real functions of real arguments. In this way, $\mathsf{P}^\mu_\nu(y)$ and $\mathsf{Q}^\mu_\nu(y)$ are Legendre functions on the cut (or Ferrer's functions) which are valid for $|y|<1$. ${\cal Q}^\mu_\nu(y)$ is the associated Legendre polynomial of the second kind and is valid for $|y|>1$. These functions are defined in App.~\ref{App:legendre} where some of the properties are also discussed (or see \cite{NIST:DLMF}). Also, we have defined for simplicity
\begin{align}
y\equiv\frac{u^2+v^2-w^{-1}}{2uv}\quad&,\quad Z^2\equiv \left(w^{-1}-(u-v)^2\right)\left((u+v)^2-w^{-1}\right)=4u^2v^2(1-y^2)\\
\tilde y=-y\quad&,\quad\tilde Z^2=-Z^2\,.
\end{align}

Using these results we can further simplify the expression for the kernel \eqref{eq:kernel2}, also expanding the Bessel functions for large argument, leading to
\begin{align}\label{eq:kernel3}
I(x\gg1,u,v)&=2^{\beta}\frac{3\sqrt{2}}{w\pi\alpha^3}\frac{1+w}{1+3w}\Gamma^2[\beta+2]{\left(uvx\right)^{-\beta-1/2}}\nonumber\\&\times\left\{\frac{\pi}{2}\sin\left(x-\frac{\beta\pi}{2}-\frac{\pi}{4}\right){I}_J(u,v,w)+\cos\left(x-\frac{\beta\pi}{2}-\frac{\pi}{4}\right){I}_Y(u,v,w)\right\}
\end{align}
where
\begin{align}\label{eq:IJJ}
I_J(u,v,w)\equiv Z^{\beta-1/2}\left[\mathsf{P}^{-\beta+1/2}_{\beta-1/2}(y)+\frac{3(1+w)}{2}\mathsf{P}^{-\beta+1/2}_{\beta+3/2}(y)\right]\Theta(u+v-w^{-1/2})
\end{align}
and
\begin{align}\label{eq:IYY}
I_Y(u,v,w)\equiv&Z^{\beta-1/2}\left[
	\mathsf{Q}^{-\beta+1/2}_{\beta-1/2}(y)+\frac{3(1+w)}{2}\mathsf{Q}^{-\beta+1/2}_{\beta+3/2}(y)\right]\Theta(u+v-w^{-1/2})\nonumber\\&
	-\tilde Z^{\beta-1/2}\left[{\cal Q}^{-\beta+1/2}_{\beta-1/2}(\tilde y)+3(1+w){\cal Q}^{-\beta+1/2}_{\beta+3/2}(\tilde y)\right]\Theta(w^{-1/2}-u-v)\,.
\end{align}
Formulas \eqref{eq:kernel3}, \eqref{eq:IJJ} and \eqref{eq:IYY} are the main result of this paper. 

\paragraph{Peaked spectrum}

Before we study a couple of particular cases, it is interesting to investigate the general behavior near the resonant point $u+v=w^{-1/2}$, which corresponds to $y=-1$. To do that, it is instructive to first look at $\mathsf{P}^\mu_\nu(y),\mathsf{Q}^\mu_\nu(y)$ near $y=-1$ where we defined
\begin{align}
\mu\equiv-\beta+1/2\quad{\rm and}\quad\nu\equiv\{\beta-1/2,\beta+3/2\}\,.
\end{align}
Note that in all cases $\mu+\nu=0,2$ and $-1<\mu\leq1/2$. In general, either (or both) functions will diverge when $y\to -1^+$. In fact, the divergence roughly goes as 
\begin{align}
\mathsf{Q}^\mu_\nu(y)&\propto \left(1+y\right)^{-|\mu|/2} \quad,\quad \mu\neq0 \quad (w\neq1/3)\,,\\
\mathsf{Q}^\mu_\nu(y)&\propto \ln(1+y) \quad,\quad \mu=0 \quad (w=1/3)\,,
\end{align}
where more details can be found in App.\ref{App:legendre}. However, the factor ${Z}\propto\sqrt{1-y^2}$ in Eqs.~\eqref{eq:IJJ} and \eqref{eq:IYY} will determine whether the kernel diverges or not in the end. Taking that factor into account we find that
\begin{align}
I(u,v,w;y\to1)&\propto \left(1+y\right)^{-\tfrac{1}{2}\left(\mu+|\mu|\right)}\quad,\quad \mu\neq0 \quad (w\neq1/3)\,,\\
I(u,v,1/3;y\to1)&\propto \ln(1+y)\quad,\quad \mu=0 \quad (w=1/3)\,.
\end{align}

On one hand, we see that indeed there is a narrow resonance for $1/3\leq w<1$ ($0\leq\mu<1/2$). This implies that for very sharp primordial power spectrum the slope of the corresponding induced GWs background will depend on the equation of state at the time of generation. It should be also noted that the equation of state can also be inferred by looking at the infrared slope of the spectrum, as pointed out in Ref.~\cite{Cai:2019cdl}. A detailed study of the infrared slope for a smooth gaussian peaked spectrum and a comparison with \cite{Cai:2019cdl} is left for future work. Now, using Eq.~\eqref{eq:spectraldensity}, the spectrum has a peak at $2\sqrt{w}k_*=1$ and grows towards the peak as
\begin{align}
\Omega_{\rm GWs}(2\sqrt{w}k\sim1)&\propto\left(1-\frac{k^2}{4wk_*^2}\right)^{\frac{2(1-3w)}{(1+3w)}}\quad,\quad 1>w>1/3\quad (1/2>\mu>0)\,,\\
\Omega_{\rm GWs}(2\sqrt{w}k\sim1)&\propto\ln^2\left(1-\frac{k^2}{4wk_*^2}\right)\quad,\quad w=1/3 \quad (\mu=0)\,.
\end{align}
The case $w=1$ ($\mu=1/2$) is special because the condition $u+v=1$ actually corresponds to $\mathbf{k}=\mathbf{q}$ (or $\theta=0$) and, therefore, there is no generation of GWs due to the fact that the polarization vector is orthogonal to the wavevector. On the other hand, we see that there is no divergence for $0<w<1/3$ ($-1<\mu<0$). Nevertheless, we numerically find that there is a peak precisely at the resonant point for all cases (see Fig.~\ref{fig:deltas}). Thus, if we assume that the primordial power spectrum has a very sharp peak at $k_*$, we can estimate the amplitude of the GWs spectrum to roughly be
\begin{align}
\Omega_{\rm GWs}(k_*,\tau)\sim {\cal E} P_{\cal R}^2\left(k_*\tau\right)^{2\frac{1-3w}{1+3w}}\,,
\end{align}
where ${\cal E}$ is an amplification factor that should be determined numerically. We also see that there is an additional enhancement or suppression depending on the value of $w$ due to the relative redshifting of GWs and the background.

\paragraph[(i)]{Scale invariant spectrum}

We can also compute the behavior of the induced SGWB for a scale invariant spectrum, \textit{i.e.} $P_{\cal R}={\rm cnt}$. In this case, the spectral density \eqref{eq:spectraldensity} reduces to
\begin{align}\label{eq:scaleinvariant}
\Omega_{\rm GWs}(x\gg1)=x^{1-2\beta}\,P_{\cal R}^2\,{\cal F}(w)\,
\end{align}
where ${\cal F}(w)$ is a numerical factor that depends on $w$ and is given by
\begin{align}
{\cal F}(w)\equiv\frac{3(1+w)^2 4^\beta}{8w^2\pi^2\alpha^6}\Gamma^4[\beta+2]\int_0^\infty dv \int_{|1-v|}^{1+v} du\left[\frac{4v^2-\left(1+v^2-u^2\right)^2}{4uv}\right]^2 \left(uv\right)^{-1-2\beta}\left(\frac{\pi^2}{4}{I}^2_J(u,v,w)+{I}^2_Y(u,v,w)\right)\,.
\end{align}
We see from Eq.~\eqref{eq:scaleinvariant} that the induced SGWB for a scale invariant scalar primordial spectrum has a scale dependence proportional to $k^{1-2\beta}$. This means that, on one hand, we recover a scale invariant spectrum for $w=1/3$ ($\beta=1/2)$. On the other hand, the spectrum will have a blue tilt for $w>1/3$ ($\beta<1/2$) and a red tilt for $0<w<1/3$ ($3/2>\beta>1/2$). In the following subsections we will consider two particular cases in more detail. We will recover the kernel for radiation domination ($w=1/3$) and we will derive new results for the case of kinetic domination ($w=1$). A particular case with $w=1/9$, where $\mu<0$, is studied in App.~\ref{app:w19} to explicitly check that the divergence does not take place.

\subsection{Radiation domination\label{subsec:RD}}

We can check that the general result Eq.~\eqref{eq:kernel3} yields the correct answer for a known case. Thus, we consider radiation domination where $w=1/3$. In that case, we have that $\beta=1/2$ ($\alpha=3/2$) and the kernel reduces to
\begin{align}
I_{RD}(x\gg 1,u,v)&=\frac{2}{uvx}\left\{-\frac{\pi}{2}\cos x \,{I}_J(u,v,1/3)+\sin x\,{I}_Y(u,v,1/3)\right\}
\end{align}
where
\begin{align}
I_J(u,v,1/3)=3y^2\Theta(u+v-\sqrt{3})
\end{align}
and
\begin{align}
I_Y(u,v,1/3)=&\left[\frac{3}{2}y\left(-2+y\ln\left(\frac{1+y}{1-y}\right)\right)
	\right]\Theta(u+v-\sqrt{3})
	\nonumber\\&-\left[\frac{3}{2}\tilde y\left(-2+\tilde y\ln\left(\frac{1+\tilde y}{1-\tilde y}\right)\right)\right]\Theta(\sqrt{3}-u-v)
	=\frac{3}{2}y\left(-2+y\ln\left|\frac{1+y}{1-y}\right|\right)\,,
\end{align}
where we have already used the particular form of the Legendre polynomials for $\beta=1/2$ that can be found in App.~\ref{App:legendre}. Joining all pieces together we find that, in the end,
\begin{align}
I_{RD}(x\gg1,u,v)=\frac{3(u^2+v^2-3)}{4u^3v^3x}&\Bigg\{-{\pi}(u^2+v^2-3)\cos x\,\Theta(u+v-\sqrt{3})\nonumber\\&+\sin x\left(-4uv+(u^2+v^2-3)\ln\left|\frac{3-(u+v)^2}{3-(u-v)^2}\right|\right)\Bigg\}\,,
\end{align}
which exactly matches the result from Kohri and Terada \cite{Kohri:2018awv}. After computing the oscillation average of the kernel squared we are led to
\begin{align}\label{eq:kernelradiation}
\overline{I_{RD}^2}(x\gg1,u,v)=\frac{9(u^2+v^2-3)^2}{32u^6v^6x^2}&\Bigg\{{\pi}^2(u^2+v^2-3)^2\Theta(u+v-\sqrt{3}) \nonumber\\&+\left(-4uv+(u^2+v^2-3)\ln\left|\frac{3-(u+v)^2}{3-(u-v)^2}\right|\right)^2\Bigg\}\,.
\end{align}

\subsection{Kinetic domination\label{subsec:KD}}
We can apply our general result to the case where the perfect fluid has a stiff equation of state ($w=1$). This stage in the early universe is common of quintessential inflation scenarios \cite{Wetterich:1987fm,Spokoiny:1993kt,Wetterich:1994bg,Peebles:1998qn,Brax:2005uf,Wetterich:2014gaa,Hossain:2014zma,Hossain:2014xha,Karananas:2016kyt,Agarwal:2017wxo,Geng:2017mic,Dimopoulos:2017zvq,Rubio:2017gty,Haro:2018zdb,GarciaBellido:2011de,Casas:2017wjh} where the inflaton fast rolls a run away potential and the universe becomes kinetic dominated. For $w=1$ we have $\beta=0$ and the kernel \eqref{eq:kernel3} takes a particular simple form since we only have to consider the region $u+v\geq 1$. We obtain for the kernel
\begin{align}
I_{KD}(x\gg 1,u,v)&=\frac{3}{\pi\sqrt{2uvx}}\left\{\frac{\pi}{2}\sin\left(x-\frac{\pi}{4}\right){I}_J(u,v,1)+\cos\left(x-\frac{\pi}{4}\right){I}_Y(u,v,1)\right\}\,,
\end{align}
where
\begin{align}
I_J(u,v,1)=\sqrt{\frac{2}{Z\pi}}\frac{3y^2-1}{(1-y^2)^{1/4}}=\frac{2}{\sqrt{\pi u v }}\frac{3y^2-1}{\sqrt{1-y^2}}
\end{align}
and
\begin{align}
I_Y(u,v,1)=-3\sqrt{\frac{2\pi}{Z}}y(1-y^2)^{1/4}=-3y\sqrt{\frac{\pi}{uv}}\,.
\end{align}
After some algebra, the kernel reads
\begin{align}
I_{KD}(x\gg1,u,v)&=\frac{3}{2\sqrt{2\pi}u^2v^2\sqrt{x}}\left\{\frac{3(u^2+v^2-1)^2-4u^2v^2}{\sqrt{4u^2v^2-(u^2+v^2-1)^2}}\sin\left(x-\frac{\pi}{4}\right)-3\left(u^2+v^2-1\right)\cos\left(x-\frac{\pi}{4}\right)\right\}\,,
\end{align}
which after taking the oscillation average of the kernel squared yields
\begin{align}\label{eq:kernelkination}
\overline{I_{KD}^2}(x\gg1,u,v)&=\frac{9}{16\pi u^4v^4x}\left\{\frac{\left(3(u^2+v^2-1)^2-4u^2v^2\right)^2}{{4u^2v^2-(u^2+v^2-1)^2}}+9\left(u^2+v^2-1\right)^2\right\}\,.
\end{align}
We are now ready to study some applications.

\begin{figure}
\centering
\includegraphics[width=0.5\textwidth]{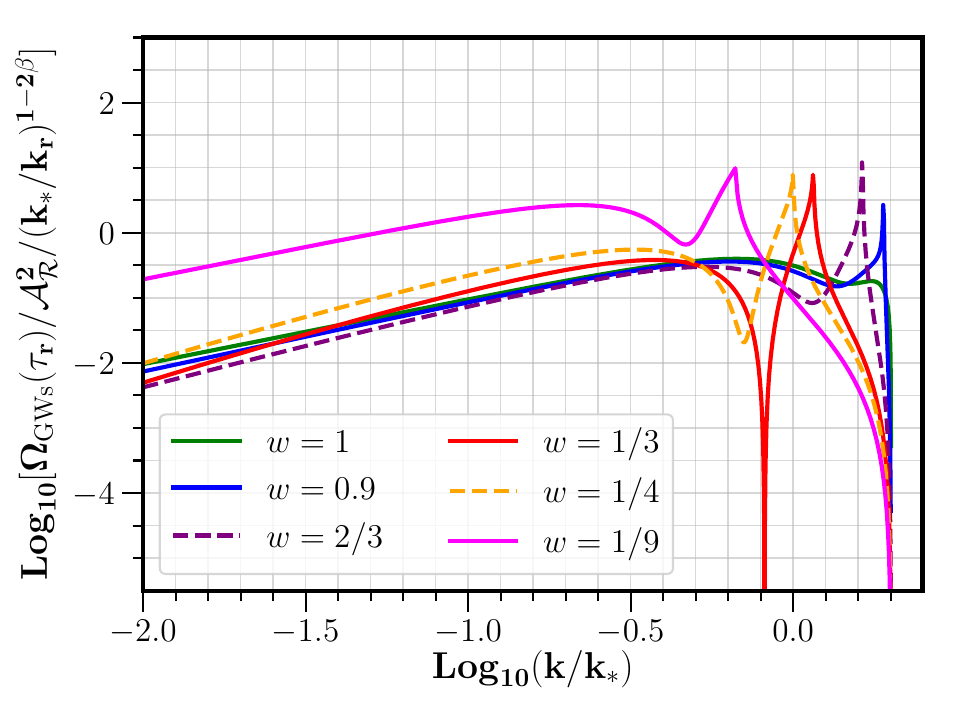}
\caption{GWs spectral density for a dirac delta power spectrum $P_{\cal R}=A_{\cal R}\delta\left(\log(k/k_*)\right)$ evaluated at a pivot time $\tau_r$ where all modes of interest are deep inside the horizon. We have divided the spectrum by the enhancement factor $\left(k_*/k_r\right)^{1-2\beta}$ for easier comparison between spectra. The corresponding pivot scale is the scale that last crossed the horizon at $\tau_r$, \textit{i.e.} $k_r\tau_r=2/(1+3w)$. We plotted the cases of $w=\{1,0.9,2/3,1/3,1/4,1/9\}$ respectively in green, blue, purple, red, orange and magenta. Although there is no divergence in the kernel for $0<w<1/3$ we see that there is a peak for all cases at the resonant scale $k=2\sqrt{w}k_*$. The only exception without peak is $w=1$ since the resonant scale is forbidden by the momentum conservation $k<2k_*$. \label{fig:deltas}}
\end{figure}

\section{Application to a peaked primordial spectrum\label{sec:PBH}}

In this section, we will be interested in the detectability of the induced SGWB and whether we can distinguish between different early universe thermal histories. We will assume that before the standard radiation domination era where the big bang nucleosynthesis (BBN) and CMB takes place, the universe was dominated by a perfect fluid with an arbitrary constant equation of state $w$. To do that, we first need to compute the spectral density of GWs today. We can use the fact that GWs decay as radiation to give the value of the spectral density of GWs during radiation domination in terms of the energy density of radiation today \cite{Kohri:2018awv}, \textit{i.e.} 
\begin{align}
\Omega_{\rm GWs}(k,\tau_0)h^2=\Omega_{\rm rad}h^2\Omega_{\rm GWs}(k,\tau_r)
\end{align}
where $\Omega_{\rm rad}h^2=4.15\times 10^{-5}$ and $\Omega_{\rm GWs}(k,\tau_r)$ has to be evaluated at the start of radiation domination era, say $\tau=\tau_r$, where (re)heating occurs. For $w=1/3$ the time $\tau_r$ is a mere reference scale for which all the modes of interest are deep inside the horizon. Also, for simplicity we will assume that (re)heating is instantaneous and leave for future work any effect of a gradual change like in Ref.~\cite{Inomata:2019zqy}. In this case, the spectral density of GWs at (re)heating reads
\begin{align}
\Omega_{\rm GWs}(k,\tau_r)=\frac{(1+3w)^2}{48}x_r^2\overline{P_h(k,\tau_r)}
\end{align}
where $x_r=k\tau_r$ and $\overline{P_h(k,\tau_r)}$ is given by Eq.~\eqref{eq:pgamma}. Under the assumptions of Sec.~\ref{sec:analyticalcalculation} we can only consider primordial scalar spectra that extend up to some scale $k_*$ well before (re)heating.

In what follows, we will use the time of instant (re)heating to set a pivot scale as the scale that last entered the horizon at $\tau_r$, that is $k_r={\cal H}_r$, which can be related to the temperature $T_r$ at (re)heating as \cite{Alabidi:2013wtp}
\begin{align}
k_r\approx 1.77\times 10^{9}{\rm Mpc}^{-1}\left(\frac{g_{*s,r}}{106.75}\right)^{1/6}\left(\frac{T_r}{10^2{\rm GeV}}\right)\,,
\end{align}
where $g_{*s,r}$ are the relativistic degrees of freedom at $T_r$. We have chosen a low heating temperature, $T_r=10^2{\rm GeV}$, so that the GW spectrum in the early phase enters the observational window. This is clear if we compute the frequency of the corresponding GWs today \cite{Alabidi:2013wtp},
\begin{align}
f\approx 2.6\times 10^{-5}{\rm Hz}\left(\frac{k}{k_r}\right)\left(\frac{g_{*s,r}}{106.75}\right)^{1/6}\left(\frac{T_r}{10^2{\rm GeV}}\right)\,,
\end{align}
where we used $f={ck}/({2\pi a})$. Now, in order to use the results of Sec.~\ref{sec:analyticalcalculation} we have to assume that the scales where the background was generated are deep inside the horizon before radiation domination starts, that is $k/k_r\gg1$. In this case, the frequency of GWs $f$ will fall inside the LISA or DECIGO windows.

\subsection{A dirac delta primordial scalar spectrum}

To have an idea of the shape of the spectrum let us consider the case of a dirac-$\delta$ type primordial spectrum \cite{Kohri:2018awv}, \textit{i.e.}
\begin{align}\label{eq:delta}
P_{\cal R}(k)={\cal A}_{\cal R}\delta\left(\ln(k/k_*)\right)\,.
\end{align}
Although this infinitely sharp spectrum is rather unphysical and a smooth gaussian might be more appropriate as in Ref.~\cite{Cai:2018dig}, it will capture the essential shape and differences in the spectra. For this simple case the power spectrum \eqref{eq:pgamma} of induced GWs reads
\begin{align}
\overline{P_h(k,\tau)}=2v_*^2{\cal A}^2_{\cal R}\left[\frac{4v_*^2-1}{4v_*^2}\right]^2 \overline{I^2}(v_*,v_*,x)\Theta\left(2v_*-1\right)\,,
\end{align}
where $v_*=k_*/k$ and the spectrum vanishes for $k>2k_*$ by momentum conservation. This result is valid for $x_r\sim k/k_r\gg1$ but this range is enough to cover the observationally interesting window.

Let us compare the two analytic results derived in the previous section. We can express the spectral density only in terms of $v_*$ (or $k/k_*$) and how far is the peak from the scale of reheating, that is $k_*/k_r$, using the fact that at horizon crossing $k_r\tau_r=2/(1+3w)$. For kinetic domination, we can use Eq.~\eqref{eq:kernelkination} evaluated at $v=u=v_*$ to find that the spectral density at reheating is given by
\begin{align}
\Omega^{KD}_{\rm GWs}(k,\tau_r)&=\frac{3{\cal A}^2_{\cal R}}{16\pi v_*^7}\left(\frac{k_*}{k_r}\right)\left(\frac{4v^2_*-1}{4v_*^2}\right)^2\left[\frac{\Big(3(2v_*^2-1)^2-4v_*^2\Big)^2}{4v_*^2-1}+9(2v_*^2-1)^2\right]\Theta\left(2v_*-1\right)\nonumber\\&=
\frac{3{\cal A}^2_{\cal R}}{64\pi v_*^{9}}\left(\frac{k_*}{k_r}\right)\left({4v^2_*-1}\right)\left[36v_*^6-60v_*^4+37v_*^2-6\right]\Theta\left(2v_*-1\right)\,.
\end{align}
For large $v_*$ ($k\ll k_*$) the spectrum falls as $k/k_*$. We also find that this spectrum has peaks at approximately $v_*\approx0.57$ and $v_*\approx1.49$ and, furthermore, is enhanced a factor $k_*/k_r$ due to the fact that the background redshifts faster than GWs. The amplitude of the spectrum at the peaks is roughly given by
\begin{align}
\Omega^{KD}_{\rm GWs}(v_*\sim 0.57)&\approx0.18\times{\cal A}^2_{\cal R}\left(\frac{k_*}{k_r}\right)
\quad {\rm and}\quad 
\Omega^{KD}_{\rm GWs}(v_*\sim 1.49)\approx0.4\times{\cal A}^2_{\cal R}\left(\frac{k_*}{k_r}\right)\,.
\end{align}

We also have to take into account the BBN bound for the energy density of GWs which can be calculated by \cite{Maggiore:1999vm,Caprini:2018mtu}
\begin{align}
\int_{k_{BBN}}^{k_{\rm end}}\Omega_{\rm GWs}h^2(k)d\ln k\leq1.12\cdot 10^{-6}\,.
\end{align}
In the kinetic domination case we find\footnote{Since BBN corresponds to $T\sim5 \,{\rm MeV}$ then $k_{BBN}/k_r\sim 10^{-6}$ and so for all practical purposes we can send $v_{BBN}=k_*/k_{BBN}\to\infty$.} that
\begin{align}
\int_{1/2}^{v_{*,BBN}}\Omega^{KD}_{\rm GWs}h^2(v_*)d\ln v_*\approx 3.4\cdot 10^{-5}{\cal A}_{\cal R}^2\left(\frac{k_*}{k_r}\right)\,.
\end{align}
Thus, we require that
\begin{align}
\frac{k_*}{k_r}<3.3\cdot 10^{-2}{\cal A}_{\cal R}^{-2}\,.
\end{align}
We see that in order to satisfy the BBN bound, the amplification factor $k_*/k_r$ is bounded by the value of the primordial scalar spectrum ${\cal A}_{\cal R}$. If $k_*/k_r$ is very large, \textit{e.g.} for very small ${\cal A}_{\cal R}$, we would go out of the observational window in frequency for future GW detectors.

\begin{figure}
\centering
\includegraphics[width=0.5\textwidth]{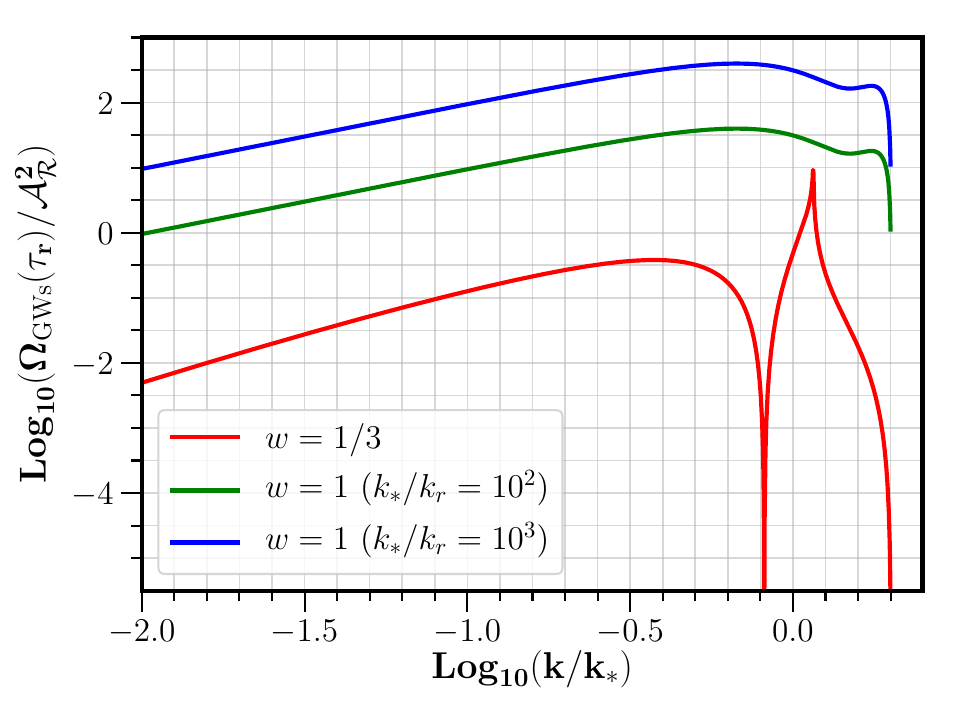}
\caption{GWs spectral density at the time of ``reheating'' for kinetic domination with $k_*/k_r=10^2,10^3$ (green and blue lines). The result for radiation domination is plotted in red. We assumed the same value $k_*$ for all cases. It should be noted that in the case of radiation domination the time of ``reheating'' should be understood as some reference time where the scales of interest are deep inside the horizon. We can see how induced GWs generated during kinetic domination are enhanced with respect to radiation domination due to the faster redshifting of the background by a factor $k_*/k_r$. We cut the spectrum at $k\sim k_r$ since the approximation $x\gg1$ breaks down there. \label{fig:delta}}
\end{figure}

Repeating a similar calculation for the radiation dominated case, we find that the spectral density is given by \cite{Kohri:2018awv}
\begin{align}
\Omega^{RD}_{\rm GWs}(k,\tau_r)=\frac{3{\cal A}^2_{\cal R}}{64v_*^{10}}\left(\frac{4v_*-1}{4v_*^2}\right)^2\left(2v_*^2-3\right)^2&\Bigg[\pi^2(2v_*^2-3)^2\Theta\left(2v_*-\sqrt{3}\right)\nonumber\\&+\left(-4v_*^2+(2v_*^2-3)\ln\left|1-\frac{4}{3}v_*^2\right|\right)^2\Bigg]\Theta\left(2v_*-1\right)\,.
\end{align}
The spectrum generated during radiation domination has a resonance at $v_*=\sqrt{3}/2$. Nevertheless, if we used a smoothed peak, \textit{e.g.} a Gaussian, the size of the peak at the resonant scale $v_*=\sqrt{3}/2$ ($k=2k_*/\sqrt{3}$) is approximately \cite{Cai:2018dig}
\begin{align}
\Omega^{RD}_{\rm GWs}(k,\tau_r,v_*\sim \sqrt{3}/2)&\approx21\times{\cal A}^2_{\cal R}\,.
\end{align}
We thus see that while the spectrum during radiation domination is enhanced due to the narrow resonance by a factor $\sim 20$, the spectrum during kinetic domination may be enhanced more by the redshifting of the background. We have plotted in Fig.~\ref{fig:delta} the induced spectral density of GWs at reheating for $w=1/3$ and for $w=1$ for two different values of $k_*/k_r$ having $k_*$ fixed. It should be noted that the case $w=1$ gives a lower bound of the power spectrum as there is no resonance. If we consider $w<1$ then there will be an additional amplification at the resonant scale.

\subsection{Primordial black hole formation}

In the last subsection we have studied the GWs signature of a peaked primordial spectrum. Interestingly, depending on the value of ${\cal A}_{\cal R}$ the density fluctuations could be large enough and collapse to PBHs with a mass spectrum roughly monochromatic \cite{Carr:2009jm,Sasaki:2018dmp}. To have an order of magnitude estimate, the mass of the PBH is approximately equal to the mass enclosed inside the horizon at the time of formation \cite{Sasaki:2018dmp}, 
\begin{align}
M_{\rm PBH}=4\pi\gamma\frac{M_{pl}^2}{H_{\rm f}}\,,
\end{align}
where $\gamma\sim0.2$ is an efficiency factor. We can relate the mass of the black hole to the pivot scale $k_r$ and the temperature of heating $T_r$ by \cite{Alabidi:2013wtp}
\begin{align}
M_{\rm PBH}\approx 1.8\times 10^{27}{\rm g}\left(\frac{k}{k_r}\right)^{-\frac{3(1+w)}{1+3w}}\left(\frac{g_{*s,r}}{106.75}\right)^{-1/2}\left(\frac{T_r}{10^2{\rm GeV}}\right)^{-2}\,.
\end{align}
The corresponding fraction of PBH at formation can be estimated to be \cite{Sasaki:2018dmp}
\begin{align}
\beta\approx\frac{\gamma}{\sqrt{2\pi}\nu_{\rm th}}e^{-\nu_{\rm th}^2/2}
\quad {\rm where}\quad
\nu_{\rm th}=\delta_{\rm th}/\sigma\quad,\quad \sigma^2=\int d\ln k W^2(kR)P_{\delta}(k)\,,
\end{align}
$W(kR)$ is a window function with scale $R$ and \cite{Harada:2013epa,Sasaki:2018dmp,Escriva:2019phb}
\begin{align}\label{eq:deltath}
\delta_{\rm th}\sim \sin^2\left(\frac{\pi\sqrt{w}}{1+3w}\right)\,.
\end{align}
However, it is important to note that the current estimate for $\delta_{\rm th}$ Eq.~\eqref{eq:deltath} is for values of $w\sim1/3$. For instance, using Eq.~\eqref{eq:deltath} we find that $\delta_{\rm th}(w=1)<\delta_{\rm th}(w=1/3)$ which cannot be correct as matter should find it harder to collapse the higher the pressure. We will take two approaches. First, we will consider a conservative estimate and use that $\delta_{\rm th}(w=1)\sim c_s^2=1$. Second, we will also use the results of \cite{Musco:2012au} where it is found that $\delta_{\rm th}(w=1/3)\sim 0.5$ and, extrapolating, $\delta_{\rm th}(w=1)\sim 0.6$. The latter values are consistent with the recent study of Ref.~\cite{Escriva:2019nsa}. It should also be noted that we are assuming gaussian primordial fluctuations and, therefore, we will not consider the role of non-gaussianities \cite{Germani:2018jgr,Atal:2018neu,Young:2019yug,Atal:2019cdz}. We leave the effects of non-gaussianities in the induced SGWB and its PBH counterpart in a general cosmological background for future work.

\begin{figure}
\centering
\includegraphics[width=0.49\textwidth]{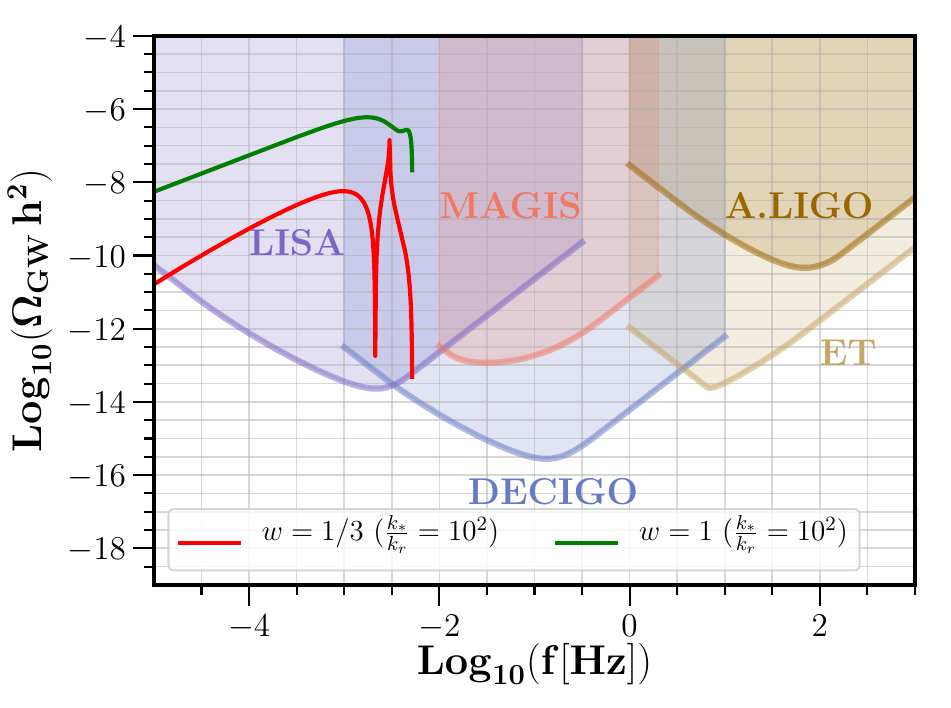}
\includegraphics[width=0.49\textwidth]{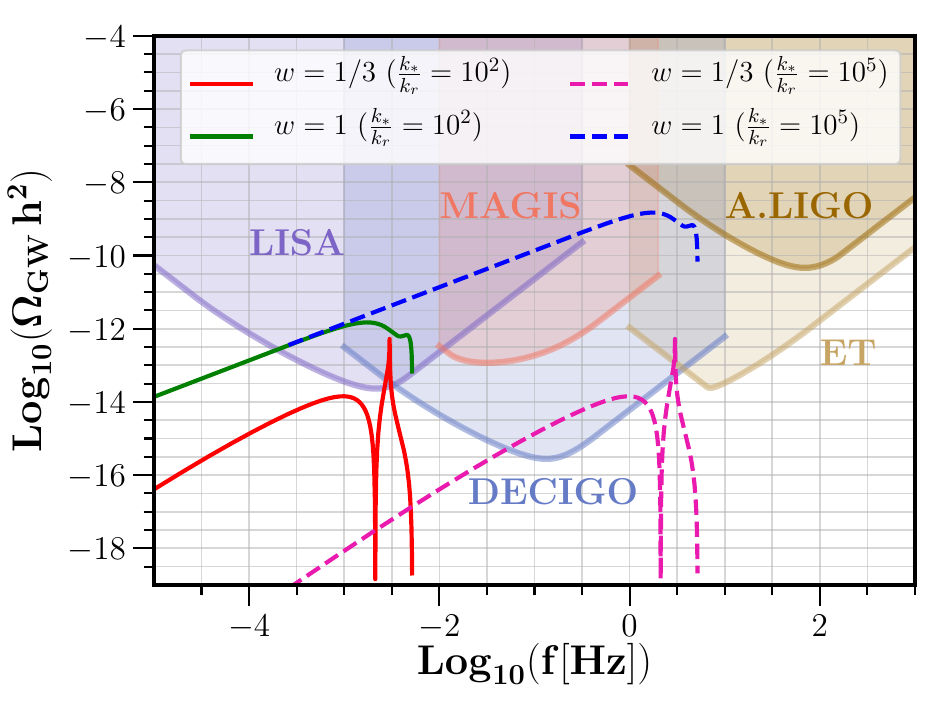}
\caption{In both plots we show the GWs spectral density today in terms of frequency. We also displayed the power-law integrated sensitivity curves \cite{Thrane:2013oya,Moore:2014lga} for LISA, DECIGO, MAGIS (space), ET and Advanced LIGO. We used a (re)heating temperature $T_r=10^2{\rm GeV}$. On the left we show the results for radiation and kinetic domination with ${\cal A}_{\cal R}=1.9\cdot10^{-2}$ and $k_*/k_r=10^2$ (respectively in red and green) that leads to $f_{\rm PBH}^{RD}=0.06$ and $f_{\rm PBH}^{KD}=0.64$ using Musco and Miller's criterion \cite{Musco:2012au}. On the right we chose ${\cal A}_{\cal R}=3\cdot10^{-5}$ for radiation and kinetic domination and $k_*/k_r=10^2$ (red and green lines) and $k_*/k_r=10^5$ (dashed blue and magenta lines). We see that induced GWs during kinetic domination are enhanced with respect to those generated during radiation domination. \label{fig:examples}}
\end{figure}

Now, using the primordial spectrum Eq.~\eqref{eq:delta} and the relation between the density contrast $\delta$ and the curvature perturbation on superhorzion scales, that is \cite{Young:2019osy}
\begin{align}
\delta=-\frac{2(1+w)}{5+3w}\left(\frac{k}{aH}\right)^2{\cal R}\,,
\end{align}
we find that
\begin{align}
\nu_{\rm th}\sim\frac{5+3w}{2(1+w)}\frac{\delta_{\rm th}}{{\cal A}_{\cal R}^{1/2}}\,,
\end{align}
where we will take $\delta_{\rm th}(w=1/3)=1/3$ (or $\delta_{\rm th}(w=1/3)=0.5$) and $\delta_{\rm th}(w=1)=1$ (or $\delta_{\rm th}(w=1)=0.6$).  Thus, for a sharp peaked primordial spectra, $\beta$ is only a function of ${\cal A}_{\cal R}$ and $w$.

Now, we can write the current fraction of PBHs with respect to dark matter by using that \cite{Sasaki:2018dmp}
\begin{align}
\beta=\frac{\rho_{\rm PBH}}{3H^2M_{pl}^2}\Bigg|_{\rm f}=\left(\frac{H_0}{H_{\rm f}}\right)^2\left(\frac{a_{\rm f}}{a_0}\right)^{-3}\Omega_{{\rm CDM}}f_{\rm PBH}\,.
\end{align}
Inverting this relation we obtain $f_{\rm PBH}$ in terms of $\beta$, \textit{i.e.}
\begin{align}\label{eq:fpbh}
f_{\rm PBH}\approx 2\beta\times 10^{15}\left(\frac{k}{k_r}\right)^{\frac{6w}{1+3w}}\left(\frac{T_r}{10^2{\rm GeV}}\right)\,,
\end{align}
where we have used that before reheating the relation between the wavenumber and the scale factor is given by
\begin{align}
\frac{k}{k_r}=\left(\frac{a_r}{a}\right)^{\tfrac{1}{2}(1+3w)}\,.
\end{align}
We then find that the fraction of PBHs \eqref{eq:fpbh} is enhanced a further $\left(\frac{k_*}{k_r}\right)^{\frac{1-3w}{1+3w}}$ with respect to the radiation domination case ($w=1/3$). 

Let us consider two particular examples. First of all, let us consider Carr's estimate \cite{Sasaki:2018dmp} $\delta_{\rm th}(w=1/3)=1/3$ and $\delta_{\rm th}(w=1)=1$. Now, let us fix that the primordial scalar spectrum has a sharp peak with amplitude ${\cal A}_{\cal R}\sim 9\cdot 10^{-3}$. On one hand, in radiation domination we have that $f^{RD}_{\rm PBH}\sim 0.54$ and $\Omega_{\rm GWs,0}^{\rm peak}\sim 7\cdot 10^{-8}$. On the other hand, in kinetic domination we instead find $f^{KD}_{\rm PBH}\sim 2\cdot 10^{-81}$ and $\Omega_{\rm GWs,0}^{\rm peak}\sim 6.8\cdot 10^{-8}$ for $k_*/k_r=10^2$. In the second example, we will consider Musco and Miller's estimate \cite{Musco:2012au} $\delta_{\rm th}(w=1/3)\sim 0.5$ and $\delta_{\rm th}(w=1)\sim 0.6$. Assuming now ${\cal A}_{\cal R}\sim 1.9\cdot 10^{-2}$, we find that, for $k_*/k_r=10^2$, $f^{RD}_{\rm PBH}\sim 0.07$ while $f^{KD}_{\rm PBH}\sim 0.64$. For $T_r\sim 10^2{\rm GeV}$ and $k_*/k_r=10^2$ these two examples correspond to $M_{\rm PBH}\sim 1.8\cdot 10^{23}g\sim 10^{-12}M_\odot$ in radiation domination and $M_{\rm PBH}\sim 1.8\cdot 10^{24}g\sim 10^{-11}M_\odot$ in kinetic domination, where $M_\odot$ is a solar mass.

We can see that, in the case at hand, the production of GWs in a kinetic dominated stage is enhanced compared to the radiation dominated one. However, the fraction of primordial black holes is very sensitive to the value of the threshold $\delta_{\rm th}$ and the scalar power spectrum ${\cal A}_{\cal R}$. For example, if we fix the value of ${\cal A}_{\cal R}$ the fraction of PBH generated during kinetic domination may be lower or higher than in radiation domination depending on whether we use Carr's or Musco and Miller's estimates respectively. A detailed numerical study like in Ref.~\cite{Escriva:2019nsa} for the case of $w\sim 1$ is needed to extract any definitive conclusion and is left for future work. Nevertheless, the difference in the value of the threshold $\delta_{\rm th}$ can be easily compensated by a higher ${\cal A}_{\cal R}$ in the kinetic dominated case. Thus, the two scenarios may only be distinguished by their induced SGWB signal. 

We plotted in Fig.~\ref{fig:examples} two examples comparing the induced SGWB generated during radiation and kinetic domination eras. First, on the left, we have plotted the above discussed case which is relevant for the PBH scenario. Second, on the right, we show two cases where the PBH production is negligible but there is a detectable induced SGWB for $k_*/k_r=10^2,10^5$. We see that the further the generation took place from (re)heating the greater the amplification in kinetic domination.

\section{Conclusions\label{sec:conclusions}}

Gravitational waves will play a crucial role in cosmology in the coming decades \cite{Maggiore:1999vm}. Particularly important will be the possible detection of a stochastic gravitational wave background as it encodes information of unexplored epochs in the universe \cite{Caprini:2018mtu}. It is thus important to consider that in the very early universe other cosmological expansions rather than radiation domination might have taken place. In this way, we studied the generation of gravitational waves at second order sourced by first order scalar squared terms, when the universe is dominated by a perfect fluid with a constant equation of state $w>0$.

We presented new analytical formulas for the kernel integrals on subhorizon scales for a general value of $w$ in Eqs.~\eqref{eq:kernel3}, \eqref{eq:IJJ} and \eqref{eq:IYY} in Sec.~\ref{sec:analyticalcalculation}. This will be a useful result for the general study of the induced stochastic gravitational wave background when $w$ is not exactly $0$ or $1/3$. Also, we found that there is a resonance in the gravitational wave spectrum at $|\mathbf{k}-\mathbf{q}|+q=k/\sqrt{w}$ when $1>w\geq1/3$. Thus, for a very sharp peaked spectrum one can find the value of $w$ by looking at the scales nearby the highest peak at $2\sqrt{w}k_*=1$. The spectrum as it approaches $k\sim k_*$ has a slope given by
\begin{align}
\ln\Omega_{\rm GWs}&\propto2\frac{1-3w}{1+3w}\ln\left(1-\frac{k^2}{4wk_*^2}\right)\quad (1>w>1/3)\,,\\ 
\ln\Omega_{\rm GWs}&\propto2\ln\left(\ln\left(1-\frac{k^2}{4wk_*^2}\right)\right) \quad w=1/3\,.
\end{align}
There is no amplification for $w=1$ as momentum conservation restricts the spectrum to $k<2k_*$. Also, no divergence of the kernel has been found for $0<w<1/3$. However, we numerically saw that there is still a peak at the resonant scale $k=2\sqrt{w}k_*$ as illustrated in Fig.~\ref{fig:deltas}. Therefore, one can infer the value of the equation of state at the time of formation by looking at the slopes of the spectrum around the resonant peak for all cases. This is a complementary result to that of Ref.~\cite{Cai:2019cdl} where it is shown that the value of $w$ can be extracted from the slope of the infrared tail of the spectrum. Then, we investigated the formula Eq.~\eqref{eq:kernel3} for two particular cases. We recovered the standard result \cite{Kohri:2018awv} for $w=1/3$ and derived new results for $w=1$ in Eq.~\eqref{eq:kernelkination} (and $w=1/9$ in App.~\ref{app:w19}).

We proceeded to study practical applications of the general result when the primordial scalar spectrum has a very sharp peak at some given scale $k_*$, which is relevant for the primordial black hole scenario \cite{Sasaki:2018dmp}. We found that in the case of kinetic domination, the spectral density of gravitational waves is enhanced by a factor $k_*/k_r\gg1$ as compared to the radiation dominated case, where $k_r$ is the last scale that entered the horizon at (re)heating. This is due to the fact that the background in a kinetic dominated universe redshifts as $a^{-6}$ while gravitational waves redshift as $a^{-4}$. 

We also studied the corresponding primordial black hole fraction in kinetic and radiation domination. However, the precise value of the primordial black hole fraction is very sensitive to the value of the threshold $\delta_{\rm th}$ and the amplitude of the scalar power spectrum ${\cal A}_{\cal R}$. Thus, a more detailed numerical study like in Ref.~\cite{Escriva:2019nsa} is needed for the case $w\sim 1$ to draw any definitive conclusion for the kinetic dominated case. For example, for a fixed value of ${\cal A}_{\cal R}$ and using Carr's criterion \cite{Sasaki:2018dmp} for the collapse we obtained that the fraction of primordial black holes created during kinetic domination, assuming $\delta_{\rm th}\sim c_s^2$ and gaussian primordial scalar fluctuations, is completely negligible than if they were created during radiation domination; despite the fact that there is also an enhancement in kinetic domination due to the faster redshifting of the background. Contrariwise, if we use Musco and Miller's results \cite{Musco:2012au} the fraction of primordial black holes in kinetic domination turns out to be higher than in radiation domination, since there is also an enhancement in the kinetic dominated case due to the faster redshifting of the background. Nevertheless, it should be noted that any difference in the value of $\delta_{\rm th}$ between $w=1/3$ and $w=1$ can be easily compensated by a higher value of ${\cal A}_{\cal R}$ in the kinetic dominated case. Therefore, the induced stochastic gravitational wave background provides a way two distinguish the PBH scenario (\textit{e.g.} given a fixed $f_{\rm PBH}$) during a cosmological expansion with general $w$. In the absence of primordial black holes, the two scenarios can also be distinguished by the shape of the gravitational waves spectrum. We leave the effects of primordial non-gaussianity for future work. 


\section*{Acknowledgments}

I would like to thank M.~Sasaki for useful feedback on the first draft and A.~D.~Rojas and D.~Rojas for insightful comments. I would also like to thank A.~Escriv\`{a}, J.~Garriga, C.~Germani, J.O.~Gong and S.~Pi for useful suggestions that improved the paper. This work was partially supported by DFG Collaborative Research center SFB 1225 (ISOQUANT)(G.D.). Calculations of cosmological perturbation theory at second order were done using the \texttt{xPand (xAct) Mathematica} package.

\appendix 

\section{Background and First order Einstein equations \label{App:first order}}
In this appendix we list the set of equations used in the main part of the paper. We divide the appendix into background and first order perturbations.

\subsection{Background equations}
Einstein equations at the background yield the Friedmann equations and the energy conservation, which are given by
\begin{align}
3{\cal H}^2&=a^2\rho\,,\\
{\cal H}'&=-\frac{1}{2}{\cal H}^2\left(1+3w\right)\,,\\
a^2\rho'&=-9\left(1+3w\right){\cal H}^2\,.
\end{align}
These equations can be solved and yield
\begin{align}
{\cal H}&=\frac{2}{\left(1+3w\right)\tau}\,.
\end{align}

\subsection{First order}
Expanding Einstein equations at first order in perturbation theory and in the Poisson gauge we find
\begin{align}
v&=-\frac{2}{3\left(1+w\right){\cal H}}\left(\Phi+\Psi'/{\cal H}\right)\,,\\
a^2c_s^2\delta\rho&=-2\left(3w{\cal H}^2\Phi-{\cal H}\Phi'-2{\cal H}\Psi'-\Psi''\right)\,,\\
\Phi&=\Psi\,.
\end{align}
They can be combined to a closed equation of motion for the gravitational potential $\Phi$ (with constant $w$),
\begin{align}
\Phi''+3(1+c_s^2){\cal H}\Phi'+3\left(c_s^2-w\right){\cal H}^2\Phi-c_s^2\Delta\Phi=0\,.
\end{align}

\section{Details on induced gravitational waves calculations \label{app:fourier}}
We use the following Fourier expansion conventions,
\begin{align}
h_{ij}=\int \frac{d^3k}{(2\pi)^3} e^{i\mathbf{k}\cdot\mathbf{x}}\left[h_R(k) {e}_{ij}^R+h_L(k) {e}_{ij}^L\right]\,,
\end{align}
where $R,L$ are the left and right polarization of gravitational waves and
\begin{align}
\Phi=\int \frac{d^3k}{(2\pi)^3} e^{i\mathbf{k}\cdot\mathbf{x}}\Phi(k)\,.
\end{align}
With this convention we can find that the power spectrum is given by
\begin{align}
\langle h_\lambda(k)h_\lambda(k')\rangle=\int_{x_0}^x d\tilde x \int_{ x_0}^x d\hat x \frac{1}{k^2{k'}^2}G(x,\tilde x)G(x,\hat x)\langle s_\lambda(k,\tilde x)s_\lambda(k',\hat x)\rangle\,.
\end{align}
Assuming Gaussianity of the primordial curvature perturbation we have that
\begin{align}
\langle s_\lambda(k,\tilde x)s_\lambda(k',\hat x)\rangle=32\int d^3q \left(e_\lambda^{ij}(k)q_iq_j\right)^2 P_\Phi(q)P_\Phi(|\mathbf{k}-\mathbf{q}|)f(q\tilde\tau,|\mathbf{k}-\mathbf{q}|\tilde\tau)f(q\hat\tau,|\mathbf{k}-\mathbf{q}|\hat\tau)\delta(k+k')\,,
\end{align}
where we used that $\langle \Phi(k)\Phi(k')\rangle=(2\pi)^3P_\Phi(k)\delta(k+k')$ and
\begin{align}
 P_\Phi(k)=\frac{2\pi^2}{k^3}\left(\frac{3+3w}{5+3w}\right)^2 \Delta_{\cal R}(k)\,.
 \end{align}
Defining
\begin{align}
\Phi(k)=\Phi_{\rm prim}(k)T(k\tau)\,,
\end{align}
the source term is given by
\begin{align}
f(q\tau,|\mathbf{k}-\mathbf{q}|\tau)=&T(q\tau)T(|\mathbf{k}-\mathbf{q}|\tau)\nonumber\\&+\frac{2}{3(1+w)}\left[T(q\tau)+\frac{1+3w}{2}\frac{dT(q\tau)}{d\ln q\tau}\right]\left[T(|\mathbf{k}-\mathbf{q}|\tau)+\frac{1+3w}{2}\frac{dT(|\mathbf{k}-\mathbf{q}|\tau)}{d\ln |\mathbf{k}-\mathbf{q}|\tau}\right]
\end{align}
and we used the fact that it is symmetric with respect to $q$ and $|\mathbf{k}-\mathbf{q}|$. 

\section{\label{app:w19} Induced gravitational waves for the special case {\texorpdfstring $aw=1/9$}}
For completeness we consider here a case with $w<1/3$ ($\mu<0$). It is convenient to study the case when $\beta=1$ which corresponds to $w=1/9$ since then we have simple expression for the Legendre polynomials. Plugging in $\beta=1$ and the values of App.~\ref{App:legendre} we find that Eq.~\eqref{eq:kernel3} reads
\begin{align}
I_{1/9}(x\gg 1,u,v)&=-\frac{45}{\pi\sqrt{2}\left(uvx\right)^{3/2}}\left\{\frac{\pi}{2}\sin\left(x+\frac{\pi}{4}\right){I}_J(u,v,1/9)+\cos\left(x+\frac{\pi}{4}\right){I}_Y(u,v,1/9)\right\}\,,
\end{align}
where
\begin{align}
I_J(u,v,1/9)&=\frac{4\sqrt{Z}}{9}\sqrt{\frac{2}{\pi}}\left(1-y^2\right)^{1/4}\left(1+5y^2\right)\Theta(u+v-\sqrt{9})\nonumber\\&
=\frac{\sqrt{4u^2v^2-(u^2+v^2-9)^2}}{9\sqrt{\pi}u^{5/2}v^{5/2}x^{3/2}}\left(4u^2v^2+5(u^2+v^2-9)^2\right)\Theta(u+v-\sqrt{9})
\end{align}
and
\begin{align}
I_Y(u,v,1/9)=&\sqrt{Z}\left[\frac{\sqrt{2\pi}}{9}\frac{10y^3-3y}{\left(1-y^2\right)^{1/4}}\right]\Theta(u+v-\sqrt{9})
	\nonumber\\&-\sqrt{\tilde Z}\left[\frac{\sqrt{2\pi}}{9}\frac{10\tilde y^3-3\tilde y-2\left(1+5\tilde y^2\right)\sqrt{\tilde y^2-1}}{\left(\tilde y^2-1\right)^{1/4}}\right]\Theta(\sqrt{9}-u-v)
	\\&
	=\frac{\sqrt{\pi}}{18u^{5/2}v^{5/2}}\left(u^2+v^2-9\right)\left[5\left(u^2+v^2-9\right)^2-6u^2v^2\right]\nonumber\\&+\frac{\sqrt{\pi}\sqrt{\left(u^2+v^2-9\right)^2-4u^2v^2}}{18u^{5/2}v^{5/2}}\left(4u^2v^2+5\left(u^2+v^2-9\right)^2\right)\Theta(\sqrt{9}-u-v)\,,
\end{align}
After some algebra, the kernel simplifies to
\begin{align}
I_{1/9}(x\gg1,u,v)&=-\frac{5}{2\sqrt{2\pi}u^4v^4{x}^{3/2}}\Bigg\{\left(u^2+v^2-9\right)\left[5\left(u^2+v^2-9\right)^2-6u^2v^2\right]\cos\left(x+\frac{\pi}{4}\right)\nonumber\\&
+\sqrt{\left|4u^2v^2-(u^2+v^2-9)^2\right|}\left(4u^2v^2+5(u^2+v^2-9)^2\right)\nonumber\\&\times \left[\sin\left(x+\frac{\pi}{4}\right)\Theta(u+v-\sqrt{9})+\cos\left(x+\frac{\pi}{4}\right)\Theta(\sqrt{9}-u-v)\right]\Bigg\}\,,
\end{align}
and after taking the oscillation average of the kernel squared yields
\begin{align}
\overline{I^2_{1/9}}(x\gg1,u,v)&=\frac{25}{16\pi u^8v^8{x}^{3}}\Bigg\{\left(u^2+v^2-9\right)^2\left[5\left(u^2+v^2-9\right)^2-6u^2v^2\right]^2\nonumber\\&
+{\left|4u^2v^2-(u^2+v^2-9)^2\right|}\left(4u^2v^2+5(u^2+v^2-9)^2\right)^2\nonumber\\&+
2\left(u^2+v^2-9\right)\left[5\left(u^2+v^2-9\right)^2-6u^2v^2\right]\sqrt{\left|4u^2v^2-(u^2+v^2-9)^2\right|}\nonumber\\&
\times\left(4u^2v^2+5(u^2+v^2-9)^2\right)\Theta(\sqrt{9}-u-v)\Bigg\}\,.
\end{align}
We see that there is no divergence arising. 

\section{Bessel functions properties \label{App:bessel}}
In this appendix we present useful properties and asymptotic behaviors for the Bessel functions.

At small arguments we have:
\begin{align}
{J_\beta(x\ll1)}\approx x^\beta\frac{2^{-\beta}}{\Gamma[1+\beta]}+O(x^{\beta+1})
\quad,\quad
{Y_\beta(x\ll1)}\approx-\frac{2^{\beta}}{\pi}{\Gamma[\beta]}x^{-\beta}+O(x^{-\beta+1})\,,
\end{align}
and for large arguments:
\begin{align}
{J_\beta(x\gg1)}\approx\sqrt{\frac{2}{\pi x}}\cos\left(x-\frac{\beta\pi}{2}-\frac{\pi}{4}\right)+O(x^{-1})
\quad ,\quad
{Y_\beta(x\gg1)}\approx\sqrt{\frac{2}{\pi x}}\sin\left(x-\frac{\beta\pi}{2}-\frac{\pi}{4}\right)+O(x^{-1})\,.
\end{align}

One can relate the derivative of a Bessel function with one lower order Bessel function as
\begin{align}
\partial_x{J}_{\nu}\left(x\right)={J}_{\nu-1}\left(x\right)-(\nu/x)%
{J}_{\nu}\left(x\right)\,,
\end{align}
and also there is a relation among the Bessel functions or nearby order as
\begin{align}
{J}_{\nu-1}\left(x\right)+{J}_{\nu+1}\left(x\right)=(2\nu/x)%
{J}_{\nu}\left(x\right)\,.
\end{align}
The Wronskian for the Green function used in the main text is given by
\begin{align}
\partial_x\left(\frac{Y_\beta(x)}{x^\beta}\right) \frac{J_\beta(x)}{x^\beta} - \partial_x\left(\frac{J_\beta(x)}{x^\beta}\right)\frac{Y_\beta(x)}{x^\beta}=\frac{2}{\pi}x^{-1-2\beta}\,.
\end{align}

\section{Analytic integrals with three Bessel functions \label{App:integralbessel}}
We review here the results of Ref.~\cite{threebesselI}. They find that for $|a-b|<c<a+b$
\begin{align}
\int_0^{\infty} d\tilde x \tilde x^{1-\beta}
\left\{	
\begin{aligned}
	J_\beta(c\tilde x)\\
	Y_\beta(c\tilde x)
\end{aligned}
\right\}
J_{\nu}(a\tilde x)J_{\nu}(b\tilde x)=\frac{1}{\pi}\sqrt{\frac{2}{\pi}}\frac{(ab)^{\beta-1}}{c^\beta}\left(\sin\varphi\right)^{\beta-1/2}\left\{	
\begin{aligned}
	\frac{\pi}{2}\mathsf{P}^{-\beta+1/2}_{\nu-1/2}(\cos\varphi)\\
	-\mathsf{Q}^{-\beta+1/2}_{\nu-1/2}(\cos\varphi)
\end{aligned}
\right\}
\end{align}
where
\begin{align}
16\Delta^2\equiv\left(c^2-(a-b)^2\right)\left((a+b)^2-c^2\right)\quad,\quad
\cos\varphi=\frac{a^2+b^2-c^2}{2ab}\quad,\quad\sin\varphi=\frac{2\Delta}{ab}\,.
\end{align}

For $c>a+b$
\begin{align}
\int_0^{\infty} d\tilde x \tilde x^{1-\beta}
&\left\{	
\begin{aligned}
	J_\beta(c\tilde x)\\
	Y_\beta(c\tilde x)
\end{aligned}
\right\}
J_{\nu}(a\tilde x)J_{\nu}(b\tilde x)\nonumber\\&
=\frac{1}{\pi}\sqrt{\frac{2}{\pi}}\frac{(ab)^{\beta-1}}{c^\beta}\left(\sinh\phi\right)^{\beta-1/2}\Gamma[\nu-\beta+1]{\cal Q}^{-\beta+1/2}_{\nu-1/2}(\cosh\phi)\left\{	
\begin{aligned}
	-\sin\left[(\nu-\beta)\pi\right]\\
	\cos\left[(\nu-\beta)\pi\right]
\end{aligned}
\right\}
\end{align}
\begin{align}
16\tilde\Delta^2\equiv\left(c^2-(a-b)^2\right)\left(c^2-(a+b)^2\right) \quad {,}\quad 
\cosh\phi=\frac{c^2-(a^2+b^2)}{2ab}\quad,\quad\sinh\phi=\frac{2\tilde\Delta}{ab}\,.
\end{align}

We can use these formulas identifying
\begin{align*}
c=1\quad,\quad a=\sqrt{w}u\quad,\quad b=\sqrt{w}v\,.
\end{align*}
In that case the range $|1-v|<u<1+v$ can be split into $1>\sqrt{w}(u+v)$ ($c>a+b$) and $1<\sqrt{w}(u+v)$ ($c<a+b$). In the latter case we always have that $\sqrt{w}|u-v|<1$ and so the formulas present above cover all range of interest.

\section{Associated Legendre functions and Legendre functions on the cut\label{App:legendre}}

We present here useful formulas that can be found in the online data base of NIST \cite{NIST:DLMF}.

The Legendre functions on the cut (or Ferrers functions) are defined for $|x|<1$ as
\begin{align}
\mathsf{P}^{\mu}_{\nu}\left(x\right)=\left(\frac{1+x}{1-x}\right)^{\mu/2}%
\mathbf{F}\left(\nu+1,-\nu;1-\mu;\tfrac{1}{2}-\tfrac{1}{2}x\right)\,,
\end{align}
\begin{align}
\mathsf{Q}^{\mu}_{\nu}\left(x\right)=\frac{\pi}{2\sin\left(\mu\pi\right)}&\Bigg\{%
\cos\left(\mu\pi\right)\left(\frac{1+x}{1-x}\right)^{\mu/2}\mathbf{F}\left(%
\nu+1,-\nu;1-\mu;\tfrac{1}{2}-\tfrac{1}{2}x\right)\\&-\frac{\Gamma\left(\nu+\mu+1%
\right)}{\Gamma\left(\nu-\mu+1\right)}\left(\frac{1-x}{1+x}\right)^{\mu/2}%
\mathbf{F}\left(\nu+1,-\nu;1+\mu;\tfrac{1}{2}-\tfrac{1}{2}x\right)\Bigg\}\,,
\end{align}
where
\begin{align}
\mathbf{F}\left(a,b;c;x\right)=\frac{1}{\Gamma\left(c\right)}F\left(a,b;c;x\right)
\end{align}
and $F\left(a,b;c;x\right)$ is the Gauss's hypergeometric function.

The associated Legendre polynomials are defined for $|x|>1$ by 
\begin{align}
P^{\mu}_{\nu}\left(x\right)=\left(\frac{x+1}{x-1}\right)^{\mu/2}\mathbf{F}%
\left(\nu+1,-\nu;1-\mu;\tfrac{1}{2}-\tfrac{1}{2}x\right)\,,
\end{align}
and
\begin{align}
Q^{\mu}_{\nu}\left(x\right)=e^{\mu\pi i}\frac{\pi^{1/2}\Gamma\left(\nu+\mu+1%
\right)\left(x^{2}-1\right)^{\mu/2}}{2^{\nu+1}x^{\nu+\mu+1}}\mathbf{F}\left(%
\tfrac{1}{2}\nu+\tfrac{1}{2}\mu+1,\tfrac{1}{2}\nu+\tfrac{1}{2}\mu+\tfrac{1}{2}%
;\nu+\tfrac{3}{2};\frac{1}{x^{2}}\right)\,.
\end{align}
It is more convenient to work with a real valuated version of the associated Legendre function of the second kind give by
\begin{align}
{\cal Q}_{\nu}^\mu(x)\equiv e^{-\mu\pi i}\frac{Q_{\nu}^\mu(x)}{\Gamma[\mu+\nu+1]}\,.
\end{align}

\subsection{Useful relations}
Some properties that help computing limits and special values of the associated Legendre polynomials and Ferrers functions are
\begin{align}
{\cal Q}_{\nu}^\mu(x)=\sqrt{\frac{\pi}{2}}\left(x^2-1\right)^{-1/4}P^{-\nu-1/2}_{-\mu-1/2}\left(x\left(x^{2}-1\right)^{-1/2}\right)\,,
\end{align}
\begin{align}
\frac{2\sin(\mu\pi)}{\pi}\mathsf{Q}_\nu^{\mu}(x)=\frac{\mathsf{P}^\mu_\nu(x)}{\Gamma[\nu+\mu+1]}-\frac{\mathsf{P}^{-\mu}_\nu(x)}{\Gamma[\nu-\mu+1]}\,,
\end{align}
\begin{align}
\mathsf{P}^{%
\mu}_{\nu}\left(-x\right)=-(2/\pi)\sin\left((\nu+\mu)\pi\right)\mathsf{Q}^{\mu}_{\nu}\left(x\right)+\cos%
\left((\nu+\mu)\pi\right)\mathsf{P}^{\mu}_{\nu}\left(x\right)\,,
\end{align}
and
\begin{align}
\mathsf{Q}^{\mu}_{\nu}\left(-x\right)=-\tfrac{1}{2}\pi\sin\left((\nu+\mu)\pi\right)\mathsf{P}^{\mu}_{\nu}\left(x%
\right)-\cos\left((\nu+\mu)\pi\right)\mathsf{Q}^{\mu}_{\nu}\left(x\right)\,.
\end{align}

For $\mu+\nu=0,2$ the reduce to
\begin{align}
\mathsf{P}^{%
\mu}_{\nu}\left(-x\right)=\mathsf{P}^{\mu}_{\nu}\left(x\right)\quad,\quad\mathsf{Q}^{\mu}_{\nu}\left(-x\right)=-\mathsf{Q}^{\mu}_{\nu}\left(x\right)\,.
\end{align}

\subsection{Singular behavior}
We present the behavior for the particular case when $\mu+\nu=0,2$ and $-1<\mu\leq1/2$ when $x\sim 1$:
\begin{align}
{\cal Q}_{\nu}^\mu(x)&\propto (x-1)^{-\mu/2} \quad \mu\neq0\,,\\
{\cal Q}_{\nu}^0(x)&\propto \ln(x-1) \quad \mu=0\,,
\end{align}
\begin{align}
\mathsf{P}_{\nu}^\mu(x)&\propto (1-x)^{-\mu/2}\,.
\end{align}

For $\mu>0$
\begin{align}
\mathsf{Q}_{\nu}^\mu(x)&\propto (1-x)^{-\mu/2} \quad \mu\neq1/2\,,\
\mathsf{Q}_{\nu}^{1/2}(x)\propto (1-x)^{1/4} \quad \mu=1/2\,,\\
\mathsf{Q}_{\nu}^0(x)&\propto \ln(1-x) \quad \mu=0\,.
\end{align}

For $\mu<0$
\begin{align}
\mathsf{Q}_{\nu}^{-|\mu|}(x)&\propto (1-x)^{-|\mu|/2}\,.
\end{align}

\subsection{Special cases}
In the main part of the text and in the appendix \ref{app:w19} we use an reduced form for the functions. We present here the forms used:
\begin{align}
\mathsf{P}_0^0(x)&=1\quad,\quad \mathsf{P}_2^0(x)=\frac{1}{2}\left(3x^2-1\right)\\
\mathsf{Q}_0^0(x)&=\frac{1}{2}\ln\left(\frac{1+x}{1-x}\right)\quad,\quad \mathsf{Q}_2^0(x)=\frac{1}{4}\left(3x^2-1\right)\ln\left(\frac{1+x}{1-x}\right)-\frac{3}{2}x\,,
\end{align}
\begin{align}
{\cal Q}_0^0(x)=\frac{1}{2}\ln\left(\frac{x+1}{x-1}\right)\quad,\quad {\cal Q}_2^0(x)=\frac{1}{8}\left(3x^2-1\right)\ln\left(\frac{x+1}{x-1}\right)-\frac{3}{4}x\,,
\end{align}
\begin{align}
\mathsf{P}^{1/2}_{-1/2}(x)&=\sqrt{\frac{2}{\pi}}\left(1-x^2\right)^{-1/4}\quad,\quad \mathsf{P}_{3/2}^{1/2}(x)=\sqrt{\frac{2}{\pi}}\frac{2x^2-1}{\left(1-x^2\right)^{1/4}}\\
\mathsf{Q}^{1/2}_{-1/2}(x)&=0\quad,\quad \mathsf{Q}_{3/2}^{1/2}(x)=-\sqrt{2\pi}x\left(1-x^2\right)^{1/4}\,,
\end{align}
\begin{align}
\mathsf{P}^{-1/2}_{1/2}(x)&=\sqrt{\frac{2}{\pi}}\left(1-x^2\right)^{1/4}\quad,\quad \mathsf{P}_{5/2}^{-1/2}(x)=\frac{1}{3}\sqrt{\frac{2}{\pi}}{\left(4x^2-1\right)}{\left(1-x^2\right)^{1/4}}\\
\mathsf{Q}^{-1/2}_{1/2}(x)&=\sqrt{\frac{\pi}{2}}\frac{x}{\left(1-x^2\right)^{1/4}}\quad,\quad \mathsf{Q}_{5/2}^{-1/2}(x)=\frac{1}{3}\sqrt{\frac{\pi}{2}}\frac{x\left(4x^2-3\right)}{\left(1-x^2\right)^{1/4}}\,,
\end{align}
\begin{align}
{\cal Q}^{-1/2}_{1/2}(x)&=\sqrt{\frac{\pi}{2}}\frac{x-\sqrt{x^2-1}}{\left(x^2-1\right)^{1/4}}\quad,\quad {\cal Q}_{5/2}^{-1/2}(x)=\frac{1}{6}\sqrt{\frac{\pi}{2}}\frac{4x^3-3x+\left(1-4x^2\right)\sqrt{x^2-1}}{\left(x^2-1\right)^{1/4}}\,.
\end{align}

\bibliography{biblio.bib}

\begin{thebibliography}{104}%
\makeatletter
\providecommand \@ifxundefined [1]{%
 \@ifx{#1\undefined}
}%
\providecommand \@ifnum [1]{%
 \ifnum #1\expandafter \@firstoftwo
 \else \expandafter \@secondoftwo
 \fi
}%
\providecommand \@ifx [1]{%
 \ifx #1\expandafter \@firstoftwo
 \else \expandafter \@secondoftwo
 \fi
}%
\providecommand \natexlab [1]{#1}%
\providecommand \enquote  [1]{``#1''}%
\providecommand \bibnamefont  [1]{#1}%
\providecommand \bibfnamefont [1]{#1}%
\providecommand \citenamefont [1]{#1}%
\providecommand \href@noop [0]{\@secondoftwo}%
\providecommand \href [0]{\begingroup \@sanitize@url \@href}%
\providecommand \@href[1]{\@@startlink{#1}\@@href}%
\providecommand \@@href[1]{\endgroup#1\@@endlink}%
\providecommand \@sanitize@url [0]{\catcode `\\12\catcode `\$12\catcode
  `\&12\catcode `\#12\catcode `\^12\catcode `\_12\catcode `\%12\relax}%
\providecommand \@@startlink[1]{}%
\providecommand \@@endlink[0]{}%
\providecommand \url  [0]{\begingroup\@sanitize@url \@url }%
\providecommand \@url [1]{\endgroup\@href {#1}{\urlprefix }}%
\providecommand \urlprefix  [0]{URL }%
\providecommand \Eprint [0]{\href }%
\providecommand \doibase [0]{http://dx.doi.org/}%
\providecommand \selectlanguage [0]{\@gobble}%
\providecommand \bibinfo  [0]{\@secondoftwo}%
\providecommand \bibfield  [0]{\@secondoftwo}%
\providecommand \translation [1]{[#1]}%
\providecommand \BibitemOpen [0]{}%
\providecommand \bibitemStop [0]{}%
\providecommand \bibitemNoStop [0]{.\EOS\space}%
\providecommand \EOS [0]{\spacefactor3000\relax}%
\providecommand \BibitemShut  [1]{\csname bibitem#1\endcsname}%
\let\auto@bib@innerbib\@empty
\bibitem [{\citenamefont {Amaro-Seoane}\ \emph {et~al.}(2017)\citenamefont
  {Amaro-Seoane} \emph {et~al.}}]{Audley:2017drz}%
  \BibitemOpen
  \bibfield  {author} {\bibinfo {author} {\bibfnamefont {P.}~\bibnamefont
  {Amaro-Seoane}} \emph {et~al.} (\bibinfo {collaboration} {LISA}),\
  }\href@noop {} {\  (\bibinfo {year} {2017})},\ \Eprint
  {http://arxiv.org/abs/1702.00786} {arXiv:1702.00786 [astro-ph.IM]}
  \BibitemShut {NoStop}%
\bibitem [{\citenamefont {Seto}\ \emph {et~al.}(2001)\citenamefont {Seto},
  \citenamefont {Kawamura},\ and\ \citenamefont {Nakamura}}]{Seto:2001qf}%
  \BibitemOpen
  \bibfield  {author} {\bibinfo {author} {\bibfnamefont {N.}~\bibnamefont
  {Seto}}, \bibinfo {author} {\bibfnamefont {S.}~\bibnamefont {Kawamura}}, \
  and\ \bibinfo {author} {\bibfnamefont {T.}~\bibnamefont {Nakamura}},\ }\href
  {\doibase 10.1103/PhysRevLett.87.221103} {\bibfield  {journal} {\bibinfo
  {journal} {Phys. Rev. Lett.}\ }\textbf {\bibinfo {volume} {87}},\ \bibinfo
  {pages} {221103} (\bibinfo {year} {2001})},\ \Eprint
  {http://arxiv.org/abs/astro-ph/0108011} {arXiv:astro-ph/0108011 [astro-ph]}
  \BibitemShut {NoStop}%
\bibitem [{\citenamefont {Yagi}\ and\ \citenamefont
  {Seto}(2011)}]{Yagi:2011wg}%
  \BibitemOpen
  \bibfield  {author} {\bibinfo {author} {\bibfnamefont {K.}~\bibnamefont
  {Yagi}}\ and\ \bibinfo {author} {\bibfnamefont {N.}~\bibnamefont {Seto}},\
  }\href {\doibase 10.1103/PhysRevD.95.109901, 10.1103/PhysRevD.83.044011}
  {\bibfield  {journal} {\bibinfo  {journal} {Phys. Rev.}\ }\textbf {\bibinfo
  {volume} {D83}},\ \bibinfo {pages} {044011} (\bibinfo {year} {2011})},\
  \bibinfo {note} {[Erratum: Phys. Rev.D95,no.10,109901(2017)]},\ \Eprint
  {http://arxiv.org/abs/1101.3940} {arXiv:1101.3940 [astro-ph.CO]} \BibitemShut
  {NoStop}%
\bibitem [{\citenamefont {Badurina}\ \emph {et~al.}(2019)\citenamefont
  {Badurina} \emph {et~al.}}]{Badurina:2019hst}%
  \BibitemOpen
  \bibfield  {author} {\bibinfo {author} {\bibfnamefont {L.}~\bibnamefont
  {Badurina}} \emph {et~al.},\ }\href@noop {} {\  (\bibinfo {year} {2019})},\
  \Eprint {http://arxiv.org/abs/1911.11755} {arXiv:1911.11755 [astro-ph.CO]}
  \BibitemShut {NoStop}%
\bibitem [{ET()}]{ET}%
  \BibitemOpen
  \href@noop {} {}\bibinfo {howpublished}
  {\url{http://www.et-gw.eu/.}}\BibitemShut {Stop}%
\bibitem [{\citenamefont {Lentati}\ \emph {et~al.}(2015)\citenamefont {Lentati}
  \emph {et~al.}}]{Lentati:2015qwp}%
  \BibitemOpen
  \bibfield  {author} {\bibinfo {author} {\bibfnamefont {L.}~\bibnamefont
  {Lentati}} \emph {et~al.},\ }\href {\doibase 10.1093/mnras/stv1538}
  {\bibfield  {journal} {\bibinfo  {journal} {Mon. Not. Roy. Astron. Soc.}\
  }\textbf {\bibinfo {volume} {453}},\ \bibinfo {pages} {2576} (\bibinfo {year}
  {2015})},\ \Eprint {http://arxiv.org/abs/1504.03692} {arXiv:1504.03692
  [astro-ph.CO]} \BibitemShut {NoStop}%
\bibitem [{\citenamefont {Shannon}\ \emph {et~al.}(2015)\citenamefont {Shannon}
  \emph {et~al.}}]{Shannon:2015ect}%
  \BibitemOpen
  \bibfield  {author} {\bibinfo {author} {\bibfnamefont {R.~M.}\ \bibnamefont
  {Shannon}} \emph {et~al.},\ }\href {\doibase 10.1126/science.aab1910}
  {\bibfield  {journal} {\bibinfo  {journal} {Science}\ }\textbf {\bibinfo
  {volume} {349}},\ \bibinfo {pages} {1522} (\bibinfo {year} {2015})},\ \Eprint
  {http://arxiv.org/abs/1509.07320} {arXiv:1509.07320 [astro-ph.CO]}
  \BibitemShut {NoStop}%
\bibitem [{\citenamefont {Arzoumanian}\ \emph {et~al.}(2016)\citenamefont
  {Arzoumanian} \emph {et~al.}}]{Arzoumanian:2015liz}%
  \BibitemOpen
  \bibfield  {author} {\bibinfo {author} {\bibfnamefont {Z.}~\bibnamefont
  {Arzoumanian}} \emph {et~al.} (\bibinfo {collaboration} {NANOGrav}),\ }\href
  {\doibase 10.3847/0004-637X/821/1/13} {\bibfield  {journal} {\bibinfo
  {journal} {Astrophys. J.}\ }\textbf {\bibinfo {volume} {821}},\ \bibinfo
  {pages} {13} (\bibinfo {year} {2016})},\ \Eprint
  {http://arxiv.org/abs/1508.03024} {arXiv:1508.03024 [astro-ph.GA]}
  \BibitemShut {NoStop}%
\bibitem [{\citenamefont {Qin}\ \emph {et~al.}(2019)\citenamefont {Qin},
  \citenamefont {Boddy}, \citenamefont {Kamionkowski},\ and\ \citenamefont
  {Dai}}]{Qin:2018yhy}%
  \BibitemOpen
  \bibfield  {author} {\bibinfo {author} {\bibfnamefont {W.}~\bibnamefont
  {Qin}}, \bibinfo {author} {\bibfnamefont {K.~K.}\ \bibnamefont {Boddy}},
  \bibinfo {author} {\bibfnamefont {M.}~\bibnamefont {Kamionkowski}}, \ and\
  \bibinfo {author} {\bibfnamefont {L.}~\bibnamefont {Dai}},\ }\href {\doibase
  10.1103/PhysRevD.99.063002} {\bibfield  {journal} {\bibinfo  {journal} {Phys.
  Rev.}\ }\textbf {\bibinfo {volume} {D99}},\ \bibinfo {pages} {063002}
  (\bibinfo {year} {2019})},\ \Eprint {http://arxiv.org/abs/1810.02369}
  {arXiv:1810.02369 [astro-ph.CO]} \BibitemShut {NoStop}%
\bibitem [{\citenamefont {Maggiore}(2000)}]{Maggiore:1999vm}%
  \BibitemOpen
  \bibfield  {author} {\bibinfo {author} {\bibfnamefont {M.}~\bibnamefont
  {Maggiore}},\ }\href {\doibase 10.1016/S0370-1573(99)00102-7} {\bibfield
  {journal} {\bibinfo  {journal} {Phys. Rept.}\ }\textbf {\bibinfo {volume}
  {331}},\ \bibinfo {pages} {283} (\bibinfo {year} {2000})},\ \Eprint
  {http://arxiv.org/abs/gr-qc/9909001} {arXiv:gr-qc/9909001 [gr-qc]}
  \BibitemShut {NoStop}%
\bibitem [{\citenamefont {Sathyaprakash}\ and\ \citenamefont
  {Schutz}(2009)}]{Sathyaprakash:2009xs}%
  \BibitemOpen
  \bibfield  {author} {\bibinfo {author} {\bibfnamefont {B.~S.}\ \bibnamefont
  {Sathyaprakash}}\ and\ \bibinfo {author} {\bibfnamefont {B.~F.}\ \bibnamefont
  {Schutz}},\ }\href {\doibase 10.12942/lrr-2009-2} {\bibfield  {journal}
  {\bibinfo  {journal} {Living Rev. Rel.}\ }\textbf {\bibinfo {volume} {12}},\
  \bibinfo {pages} {2} (\bibinfo {year} {2009})},\ \Eprint
  {http://arxiv.org/abs/0903.0338} {arXiv:0903.0338 [gr-qc]} \BibitemShut
  {NoStop}%
\bibitem [{\citenamefont {Guzzetti}\ \emph {et~al.}(2016)\citenamefont
  {Guzzetti}, \citenamefont {Bartolo}, \citenamefont {Liguori},\ and\
  \citenamefont {Matarrese}}]{Guzzetti:2016mkm}%
  \BibitemOpen
  \bibfield  {author} {\bibinfo {author} {\bibfnamefont {M.~C.}\ \bibnamefont
  {Guzzetti}}, \bibinfo {author} {\bibfnamefont {N.}~\bibnamefont {Bartolo}},
  \bibinfo {author} {\bibfnamefont {M.}~\bibnamefont {Liguori}}, \ and\
  \bibinfo {author} {\bibfnamefont {S.}~\bibnamefont {Matarrese}},\ }\href
  {\doibase 10.1393/ncr/i2016-10127-1} {\bibfield  {journal} {\bibinfo
  {journal} {Riv. Nuovo Cim.}\ }\textbf {\bibinfo {volume} {39}},\ \bibinfo
  {pages} {399} (\bibinfo {year} {2016})},\ \Eprint
  {http://arxiv.org/abs/1605.01615} {arXiv:1605.01615 [astro-ph.CO]}
  \BibitemShut {NoStop}%
\bibitem [{\citenamefont {{Caprini}}\ and\ \citenamefont
  {Figueroa}(2018)}]{Caprini:2018mtu}%
  \BibitemOpen
  \bibfield  {author} {\bibinfo {author} {\bibfnamefont {C.}~\bibnamefont
  {{Caprini}}}\ and\ \bibinfo {author} {\bibfnamefont {D.~G.}\ \bibnamefont
  {Figueroa}},\ }\href {\doibase 10.1088/1361-6382/aac608} {\bibfield
  {journal} {\bibinfo  {journal} {Class. Quant. Grav.}\ }\textbf {\bibinfo
  {volume} {35}},\ \bibinfo {pages} {163001} (\bibinfo {year} {2018})},\
  \Eprint {http://arxiv.org/abs/1801.04268} {arXiv:1801.04268 [astro-ph.CO]}
  \BibitemShut {NoStop}%
\bibitem [{\citenamefont {Christensen}(2019)}]{Christensen:2018iqi}%
  \BibitemOpen
  \bibfield  {author} {\bibinfo {author} {\bibfnamefont {N.}~\bibnamefont
  {Christensen}},\ }\href {\doibase 10.1088/1361-6633/aae6b5} {\bibfield
  {journal} {\bibinfo  {journal} {Rept. Prog. Phys.}\ }\textbf {\bibinfo
  {volume} {82}},\ \bibinfo {pages} {016903} (\bibinfo {year} {2019})},\
  \Eprint {http://arxiv.org/abs/1811.08797} {arXiv:1811.08797 [gr-qc]}
  \BibitemShut {NoStop}%
\bibitem [{\citenamefont {Regimbau}(2011)}]{Regimbau:2011rp}%
  \BibitemOpen
  \bibfield  {author} {\bibinfo {author} {\bibfnamefont {T.}~\bibnamefont
  {Regimbau}},\ }\href {\doibase 10.1088/1674-4527/11/4/001} {\bibfield
  {journal} {\bibinfo  {journal} {Res. Astron. Astrophys.}\ }\textbf {\bibinfo
  {volume} {11}},\ \bibinfo {pages} {369} (\bibinfo {year} {2011})},\ \Eprint
  {http://arxiv.org/abs/1101.2762} {arXiv:1101.2762 [astro-ph.CO]} \BibitemShut
  {NoStop}%
\bibitem [{\citenamefont {Akrami}\ \emph {et~al.}(2018)\citenamefont {Akrami}
  \emph {et~al.}}]{Akrami:2018odb}%
  \BibitemOpen
  \bibfield  {author} {\bibinfo {author} {\bibfnamefont {Y.}~\bibnamefont
  {Akrami}} \emph {et~al.} (\bibinfo {collaboration} {Planck}),\ }\href@noop {}
  {\  (\bibinfo {year} {2018})},\ \Eprint {http://arxiv.org/abs/1807.06211}
  {arXiv:1807.06211 [astro-ph.CO]} \BibitemShut {NoStop}%
\bibitem [{\citenamefont {Aghanim}\ \emph {et~al.}(2018)\citenamefont {Aghanim}
  \emph {et~al.}}]{Aghanim:2018eyx}%
  \BibitemOpen
  \bibfield  {author} {\bibinfo {author} {\bibfnamefont {N.}~\bibnamefont
  {Aghanim}} \emph {et~al.} (\bibinfo {collaboration} {Planck}),\ }\href@noop
  {} {\  (\bibinfo {year} {2018})},\ \Eprint {http://arxiv.org/abs/1807.06209}
  {arXiv:1807.06209 [astro-ph.CO]} \BibitemShut {NoStop}%
\bibitem [{\citenamefont {Brout}\ \emph {et~al.}(1978)\citenamefont {Brout},
  \citenamefont {Englert},\ and\ \citenamefont {Gunzig}}]{Brout:1977ix}%
  \BibitemOpen
  \bibfield  {author} {\bibinfo {author} {\bibfnamefont {R.}~\bibnamefont
  {Brout}}, \bibinfo {author} {\bibfnamefont {F.}~\bibnamefont {Englert}}, \
  and\ \bibinfo {author} {\bibfnamefont {E.}~\bibnamefont {Gunzig}},\ }\href
  {\doibase 10.1016/0003-4916(78)90176-8} {\bibfield  {journal} {\bibinfo
  {journal} {Annals Phys.}\ }\textbf {\bibinfo {volume} {115}},\ \bibinfo
  {pages} {78} (\bibinfo {year} {1978})}\BibitemShut {NoStop}%
\bibitem [{\citenamefont {Starobinsky}(1979)}]{Starobinsky:1979ty}%
  \BibitemOpen
  \bibfield  {author} {\bibinfo {author} {\bibfnamefont {A.~A.}\ \bibnamefont
  {Starobinsky}},\ }\href@noop {} {\bibfield  {journal} {\bibinfo  {journal}
  {JETP Lett.}\ }\textbf {\bibinfo {volume} {30}},\ \bibinfo {pages} {682}
  (\bibinfo {year} {1979})}\BibitemShut {NoStop}%
\bibitem [{\citenamefont {Guth}(1981)}]{Guth:1980zm}%
  \BibitemOpen
  \bibfield  {author} {\bibinfo {author} {\bibfnamefont {A.~H.}\ \bibnamefont
  {Guth}},\ }\href {\doibase 10.1103/PhysRevD.23.347} {\bibfield  {journal}
  {\bibinfo  {journal} {Phys. Rev.}\ }\textbf {\bibinfo {volume} {D23}},\
  \bibinfo {pages} {347} (\bibinfo {year} {1981})}\BibitemShut {NoStop}%
\bibitem [{\citenamefont {Sato}(1981)}]{Sato:1980yn}%
  \BibitemOpen
  \bibfield  {author} {\bibinfo {author} {\bibfnamefont {K.}~\bibnamefont
  {Sato}},\ }\href@noop {} {\bibfield  {journal} {\bibinfo  {journal} {Mon.
  Not. Roy. Astron. Soc.}\ }\textbf {\bibinfo {volume} {195}},\ \bibinfo
  {pages} {467} (\bibinfo {year} {1981})}\BibitemShut {NoStop}%
\bibitem [{\citenamefont {Assadullahi}\ and\ \citenamefont
  {Wands}(2010)}]{Assadullahi:2009jc}%
  \BibitemOpen
  \bibfield  {author} {\bibinfo {author} {\bibfnamefont {H.}~\bibnamefont
  {Assadullahi}}\ and\ \bibinfo {author} {\bibfnamefont {D.}~\bibnamefont
  {Wands}},\ }\href {\doibase 10.1103/PhysRevD.81.023527} {\bibfield  {journal}
  {\bibinfo  {journal} {Phys. Rev.}\ }\textbf {\bibinfo {volume} {D81}},\
  \bibinfo {pages} {023527} (\bibinfo {year} {2010})},\ \Eprint
  {http://arxiv.org/abs/0907.4073} {arXiv:0907.4073 [astro-ph.CO]} \BibitemShut
  {NoStop}%
\bibitem [{\citenamefont {Bugaev}\ and\ \citenamefont
  {Klimai}(2011)}]{Bugaev:2010bb}%
  \BibitemOpen
  \bibfield  {author} {\bibinfo {author} {\bibfnamefont {E.}~\bibnamefont
  {Bugaev}}\ and\ \bibinfo {author} {\bibfnamefont {P.}~\bibnamefont
  {Klimai}},\ }\href {\doibase 10.1103/PhysRevD.83.083521} {\bibfield
  {journal} {\bibinfo  {journal} {Phys. Rev.}\ }\textbf {\bibinfo {volume}
  {D83}},\ \bibinfo {pages} {083521} (\bibinfo {year} {2011})},\ \Eprint
  {http://arxiv.org/abs/1012.4697} {arXiv:1012.4697 [astro-ph.CO]} \BibitemShut
  {NoStop}%
\bibitem [{\citenamefont {Inomata}\ and\ \citenamefont
  {Nakama}(2019)}]{Inomata:2018epa}%
  \BibitemOpen
  \bibfield  {author} {\bibinfo {author} {\bibfnamefont {K.}~\bibnamefont
  {Inomata}}\ and\ \bibinfo {author} {\bibfnamefont {T.}~\bibnamefont
  {Nakama}},\ }\href {\doibase 10.1103/PhysRevD.99.043511} {\bibfield
  {journal} {\bibinfo  {journal} {Phys. Rev.}\ }\textbf {\bibinfo {volume}
  {D99}},\ \bibinfo {pages} {043511} (\bibinfo {year} {2019})},\ \Eprint
  {http://arxiv.org/abs/1812.00674} {arXiv:1812.00674 [astro-ph.CO]}
  \BibitemShut {NoStop}%
\bibitem [{\citenamefont {Sato-Polito}\ \emph {et~al.}(2019)\citenamefont
  {Sato-Polito}, \citenamefont {Kovetz},\ and\ \citenamefont
  {Kamionkowski}}]{Sato-Polito:2019hws}%
  \BibitemOpen
  \bibfield  {author} {\bibinfo {author} {\bibfnamefont {G.}~\bibnamefont
  {Sato-Polito}}, \bibinfo {author} {\bibfnamefont {E.~D.}\ \bibnamefont
  {Kovetz}}, \ and\ \bibinfo {author} {\bibfnamefont {M.}~\bibnamefont
  {Kamionkowski}},\ }\href {\doibase 10.1103/PhysRevD.100.063521} {\bibfield
  {journal} {\bibinfo  {journal} {Phys. Rev.}\ }\textbf {\bibinfo {volume}
  {D100}},\ \bibinfo {pages} {063521} (\bibinfo {year} {2019})},\ \Eprint
  {http://arxiv.org/abs/1904.10971} {arXiv:1904.10971 [astro-ph.CO]}
  \BibitemShut {NoStop}%
\bibitem [{\citenamefont {Kodama}\ and\ \citenamefont
  {{Sasaki}}(1984)}]{Kodama:1985bj}%
  \BibitemOpen
  \bibfield  {author} {\bibinfo {author} {\bibfnamefont {H.}~\bibnamefont
  {Kodama}}\ and\ \bibinfo {author} {\bibfnamefont {M.}~\bibnamefont
  {{Sasaki}}},\ }\href {\doibase 10.1143/PTPS.78.1} {\bibfield  {journal}
  {\bibinfo  {journal} {Prog. Theor. Phys. Suppl.}\ }\textbf {\bibinfo {volume}
  {78}},\ \bibinfo {pages} {1} (\bibinfo {year} {1984})}\BibitemShut {NoStop}%
\bibitem [{\citenamefont {Mukhanov}\ \emph {et~al.}(1992)\citenamefont
  {Mukhanov}, \citenamefont {Feldman},\ and\ \citenamefont
  {Brandenberger}}]{Mukhanov:1990me}%
  \BibitemOpen
  \bibfield  {author} {\bibinfo {author} {\bibfnamefont {V.~F.}\ \bibnamefont
  {Mukhanov}}, \bibinfo {author} {\bibfnamefont {H.~A.}\ \bibnamefont
  {Feldman}}, \ and\ \bibinfo {author} {\bibfnamefont {R.~H.}\ \bibnamefont
  {Brandenberger}},\ }\href {\doibase 10.1016/0370-1573(92)90044-Z} {\bibfield
  {journal} {\bibinfo  {journal} {Phys. Rept.}\ }\textbf {\bibinfo {volume}
  {215}},\ \bibinfo {pages} {203} (\bibinfo {year} {1992})}\BibitemShut
  {NoStop}%
\bibitem [{\citenamefont {Noh}\ and\ \citenamefont {Hwang}(2004)}]{Noh:2004bc}%
  \BibitemOpen
  \bibfield  {author} {\bibinfo {author} {\bibfnamefont {H.}~\bibnamefont
  {Noh}}\ and\ \bibinfo {author} {\bibfnamefont {J.-c.}\ \bibnamefont
  {Hwang}},\ }\href {\doibase 10.1103/PhysRevD.69.104011} {\bibfield  {journal}
  {\bibinfo  {journal} {Phys. Rev.}\ }\textbf {\bibinfo {volume} {D69}},\
  \bibinfo {pages} {104011} (\bibinfo {year} {2004})}\BibitemShut {NoStop}%
\bibitem [{\citenamefont {Hwang}\ and\ \citenamefont
  {Noh}(2007)}]{Hwang:2007ni}%
  \BibitemOpen
  \bibfield  {author} {\bibinfo {author} {\bibfnamefont {J.-c.}\ \bibnamefont
  {Hwang}}\ and\ \bibinfo {author} {\bibfnamefont {H.}~\bibnamefont {Noh}},\
  }\href {\doibase 10.1103/PhysRevD.76.103527} {\bibfield  {journal} {\bibinfo
  {journal} {Phys. Rev.}\ }\textbf {\bibinfo {volume} {D76}},\ \bibinfo {pages}
  {103527} (\bibinfo {year} {2007})},\ \Eprint {http://arxiv.org/abs/0704.1927}
  {arXiv:0704.1927 [astro-ph]} \BibitemShut {NoStop}%
\bibitem [{\citenamefont {Ananda}\ \emph {et~al.}(2007)\citenamefont {Ananda},
  \citenamefont {Clarkson},\ and\ \citenamefont {Wands}}]{Ananda:2006af}%
  \BibitemOpen
  \bibfield  {author} {\bibinfo {author} {\bibfnamefont {K.~N.}\ \bibnamefont
  {Ananda}}, \bibinfo {author} {\bibfnamefont {C.}~\bibnamefont {Clarkson}}, \
  and\ \bibinfo {author} {\bibfnamefont {D.}~\bibnamefont {Wands}},\ }\href
  {\doibase 10.1103/PhysRevD.75.123518} {\bibfield  {journal} {\bibinfo
  {journal} {Phys. Rev.}\ }\textbf {\bibinfo {volume} {D75}},\ \bibinfo {pages}
  {123518} (\bibinfo {year} {2007})},\ \Eprint
  {http://arxiv.org/abs/gr-qc/0612013} {arXiv:gr-qc/0612013 [gr-qc]}
  \BibitemShut {NoStop}%
\bibitem [{\citenamefont {Baumann}\ \emph {et~al.}(2007)\citenamefont
  {Baumann}, \citenamefont {Steinhardt}, \citenamefont {Takahashi},\ and\
  \citenamefont {Ichiki}}]{Baumann:2007zm}%
  \BibitemOpen
  \bibfield  {author} {\bibinfo {author} {\bibfnamefont {D.}~\bibnamefont
  {Baumann}}, \bibinfo {author} {\bibfnamefont {P.~J.}\ \bibnamefont
  {Steinhardt}}, \bibinfo {author} {\bibfnamefont {K.}~\bibnamefont
  {Takahashi}}, \ and\ \bibinfo {author} {\bibfnamefont {K.}~\bibnamefont
  {Ichiki}},\ }\href {\doibase 10.1103/PhysRevD.76.084019} {\bibfield
  {journal} {\bibinfo  {journal} {Phys. Rev.}\ }\textbf {\bibinfo {volume}
  {D76}},\ \bibinfo {pages} {084019} (\bibinfo {year} {2007})},\ \Eprint
  {http://arxiv.org/abs/hep-th/0703290} {arXiv:hep-th/0703290 [hep-th]}
  \BibitemShut {NoStop}%
\bibitem [{\citenamefont {Alabidi}\ \emph {et~al.}(2012)\citenamefont
  {Alabidi}, \citenamefont {Kohri}, \citenamefont {Sasaki},\ and\ \citenamefont
  {Sendouda}}]{Alabidi:2012ex}%
  \BibitemOpen
  \bibfield  {author} {\bibinfo {author} {\bibfnamefont {L.}~\bibnamefont
  {Alabidi}}, \bibinfo {author} {\bibfnamefont {K.}~\bibnamefont {Kohri}},
  \bibinfo {author} {\bibfnamefont {M.}~\bibnamefont {Sasaki}}, \ and\ \bibinfo
  {author} {\bibfnamefont {Y.}~\bibnamefont {Sendouda}},\ }\href {\doibase
  10.1088/1475-7516/2012/09/017} {\bibfield  {journal} {\bibinfo  {journal}
  {JCAP}\ }\textbf {\bibinfo {volume} {1209}},\ \bibinfo {pages} {017}
  (\bibinfo {year} {2012})},\ \Eprint {http://arxiv.org/abs/1203.4663}
  {arXiv:1203.4663 [astro-ph.CO]} \BibitemShut {NoStop}%
\bibitem [{\citenamefont {Alabidi}\ \emph {et~al.}(2013)\citenamefont
  {Alabidi}, \citenamefont {Kohri}, \citenamefont {{Sasaki}},\ and\
  \citenamefont {Sendouda}}]{Alabidi:2013wtp}%
  \BibitemOpen
  \bibfield  {author} {\bibinfo {author} {\bibfnamefont {L.}~\bibnamefont
  {Alabidi}}, \bibinfo {author} {\bibfnamefont {K.}~\bibnamefont {Kohri}},
  \bibinfo {author} {\bibfnamefont {M.}~\bibnamefont {{Sasaki}}}, \ and\
  \bibinfo {author} {\bibfnamefont {Y.}~\bibnamefont {Sendouda}},\ }\href
  {\doibase 10.1088/1475-7516/2013/05/033} {\bibfield  {journal} {\bibinfo
  {journal} {JCAP}\ }\textbf {\bibinfo {volume} {1305}},\ \bibinfo {pages}
  {033} (\bibinfo {year} {2013})},\ \Eprint {http://arxiv.org/abs/1303.4519}
  {arXiv:1303.4519 [astro-ph.CO]} \BibitemShut {NoStop}%
\bibitem [{\citenamefont {Hwang}\ \emph {et~al.}(2017)\citenamefont {Hwang},
  \citenamefont {Jeong},\ and\ \citenamefont {Noh}}]{Hwang:2017oxa}%
  \BibitemOpen
  \bibfield  {author} {\bibinfo {author} {\bibfnamefont {J.-C.}\ \bibnamefont
  {Hwang}}, \bibinfo {author} {\bibfnamefont {D.}~\bibnamefont {Jeong}}, \ and\
  \bibinfo {author} {\bibfnamefont {H.}~\bibnamefont {Noh}},\ }\href {\doibase
  10.3847/1538-4357/aa74be} {\bibfield  {journal} {\bibinfo  {journal}
  {Astrophys. J.}\ }\textbf {\bibinfo {volume} {842}},\ \bibinfo {pages} {46}
  (\bibinfo {year} {2017})},\ \Eprint {http://arxiv.org/abs/1704.03500}
  {arXiv:1704.03500 [astro-ph.CO]} \BibitemShut {NoStop}%
\bibitem [{\citenamefont {Kohri}\ and\ \citenamefont
  {Terada}(2018)}]{Kohri:2018awv}%
  \BibitemOpen
  \bibfield  {author} {\bibinfo {author} {\bibfnamefont {K.}~\bibnamefont
  {Kohri}}\ and\ \bibinfo {author} {\bibfnamefont {T.}~\bibnamefont {Terada}},\
  }\href {\doibase 10.1103/PhysRevD.97.123532} {\bibfield  {journal} {\bibinfo
  {journal} {Phys. Rev.}\ }\textbf {\bibinfo {volume} {D97}},\ \bibinfo {pages}
  {123532} (\bibinfo {year} {2018})},\ \Eprint
  {http://arxiv.org/abs/1804.08577} {arXiv:1804.08577 [gr-qc]} \BibitemShut
  {NoStop}%
\bibitem [{\citenamefont {Cai}\ \emph {et~al.}(2019{\natexlab{a}})\citenamefont
  {Cai}, \citenamefont {Pi},\ and\ \citenamefont {Sasaki}}]{Cai:2018dig}%
  \BibitemOpen
  \bibfield  {author} {\bibinfo {author} {\bibfnamefont {R.-g.}\ \bibnamefont
  {Cai}}, \bibinfo {author} {\bibfnamefont {S.}~\bibnamefont {Pi}}, \ and\
  \bibinfo {author} {\bibfnamefont {M.}~\bibnamefont {Sasaki}},\ }\href
  {\doibase 10.1103/PhysRevLett.122.201101} {\bibfield  {journal} {\bibinfo
  {journal} {Phys. Rev. Lett.}\ }\textbf {\bibinfo {volume} {122}},\ \bibinfo
  {pages} {201101} (\bibinfo {year} {2019}{\natexlab{a}})},\ \Eprint
  {http://arxiv.org/abs/1810.11000} {arXiv:1810.11000 [astro-ph.CO]}
  \BibitemShut {NoStop}%
\bibitem [{\citenamefont {Bartolo}\ \emph {et~al.}(2019)\citenamefont
  {Bartolo}, \citenamefont {De~Luca}, \citenamefont {Franciolini},
  \citenamefont {Peloso}, \citenamefont {Racco},\ and\ \citenamefont
  {Riotto}}]{Bartolo:2018rku}%
  \BibitemOpen
  \bibfield  {author} {\bibinfo {author} {\bibfnamefont {N.}~\bibnamefont
  {Bartolo}}, \bibinfo {author} {\bibfnamefont {V.}~\bibnamefont {De~Luca}},
  \bibinfo {author} {\bibfnamefont {G.}~\bibnamefont {Franciolini}}, \bibinfo
  {author} {\bibfnamefont {M.}~\bibnamefont {Peloso}}, \bibinfo {author}
  {\bibfnamefont {D.}~\bibnamefont {Racco}}, \ and\ \bibinfo {author}
  {\bibfnamefont {A.}~\bibnamefont {Riotto}},\ }\href {\doibase
  10.1103/PhysRevD.99.103521} {\bibfield  {journal} {\bibinfo  {journal} {Phys.
  Rev.}\ }\textbf {\bibinfo {volume} {D99}},\ \bibinfo {pages} {103521}
  (\bibinfo {year} {2019})},\ \Eprint {http://arxiv.org/abs/1810.12224}
  {arXiv:1810.12224 [astro-ph.CO]} \BibitemShut {NoStop}%
\bibitem [{\citenamefont {Yuan}\ \emph
  {et~al.}(2019{\natexlab{a}})\citenamefont {Yuan}, \citenamefont {Chen},\ and\
  \citenamefont {Huang}}]{Yuan:2019udt}%
  \BibitemOpen
  \bibfield  {author} {\bibinfo {author} {\bibfnamefont {C.}~\bibnamefont
  {Yuan}}, \bibinfo {author} {\bibfnamefont {Z.-C.}\ \bibnamefont {Chen}}, \
  and\ \bibinfo {author} {\bibfnamefont {Q.-G.}\ \bibnamefont {Huang}},\ }\href
  {\doibase 10.1103/PhysRevD.100.081301} {\bibfield  {journal} {\bibinfo
  {journal} {Phys. Rev.}\ }\textbf {\bibinfo {volume} {D100}},\ \bibinfo
  {pages} {081301} (\bibinfo {year} {2019}{\natexlab{a}})},\ \Eprint
  {http://arxiv.org/abs/1906.11549} {arXiv:1906.11549 [astro-ph.CO]}
  \BibitemShut {NoStop}%
\bibitem [{\citenamefont {Inomata}\ \emph
  {et~al.}(2019{\natexlab{a}})\citenamefont {Inomata}, \citenamefont {Kohri},
  \citenamefont {Nakama},\ and\ \citenamefont {Terada}}]{Inomata:2019zqy}%
  \BibitemOpen
  \bibfield  {author} {\bibinfo {author} {\bibfnamefont {K.}~\bibnamefont
  {Inomata}}, \bibinfo {author} {\bibfnamefont {K.}~\bibnamefont {Kohri}},
  \bibinfo {author} {\bibfnamefont {T.}~\bibnamefont {Nakama}}, \ and\ \bibinfo
  {author} {\bibfnamefont {T.}~\bibnamefont {Terada}},\ }\href {\doibase
  10.1088/1475-7516/2019/10/071} {\bibfield  {journal} {\bibinfo  {journal}
  {JCAP}\ }\textbf {\bibinfo {volume} {1910}},\ \bibinfo {pages} {071}
  (\bibinfo {year} {2019}{\natexlab{a}})},\ \Eprint
  {http://arxiv.org/abs/1904.12878} {arXiv:1904.12878 [astro-ph.CO]}
  \BibitemShut {NoStop}%
\bibitem [{\citenamefont {Inomata}\ \emph
  {et~al.}(2019{\natexlab{b}})\citenamefont {Inomata}, \citenamefont {Kohri},
  \citenamefont {Nakama},\ and\ \citenamefont {Terada}}]{Inomata:2019ivs}%
  \BibitemOpen
  \bibfield  {author} {\bibinfo {author} {\bibfnamefont {K.}~\bibnamefont
  {Inomata}}, \bibinfo {author} {\bibfnamefont {K.}~\bibnamefont {Kohri}},
  \bibinfo {author} {\bibfnamefont {T.}~\bibnamefont {Nakama}}, \ and\ \bibinfo
  {author} {\bibfnamefont {T.}~\bibnamefont {Terada}},\ }\href@noop {} {\
  (\bibinfo {year} {2019}{\natexlab{b}})},\ \Eprint
  {http://arxiv.org/abs/1904.12879} {arXiv:1904.12879 [astro-ph.CO]}
  \BibitemShut {NoStop}%
\bibitem [{\citenamefont {Chen}\ \emph {et~al.}(2019)\citenamefont {Chen},
  \citenamefont {Yuan},\ and\ \citenamefont {Huang}}]{Chen:2019xse}%
  \BibitemOpen
  \bibfield  {author} {\bibinfo {author} {\bibfnamefont {Z.-C.}\ \bibnamefont
  {Chen}}, \bibinfo {author} {\bibfnamefont {C.}~\bibnamefont {Yuan}}, \ and\
  \bibinfo {author} {\bibfnamefont {Q.-G.}\ \bibnamefont {Huang}},\ }\href@noop
  {} {\  (\bibinfo {year} {2019})},\ \Eprint {http://arxiv.org/abs/1910.12239}
  {arXiv:1910.12239 [astro-ph.CO]} \BibitemShut {NoStop}%
\bibitem [{\citenamefont {Yuan}\ \emph
  {et~al.}(2019{\natexlab{b}})\citenamefont {Yuan}, \citenamefont {Chen},\ and\
  \citenamefont {Huang}}]{Yuan:2019wwo}%
  \BibitemOpen
  \bibfield  {author} {\bibinfo {author} {\bibfnamefont {C.}~\bibnamefont
  {Yuan}}, \bibinfo {author} {\bibfnamefont {Z.-C.}\ \bibnamefont {Chen}}, \
  and\ \bibinfo {author} {\bibfnamefont {Q.-G.}\ \bibnamefont {Huang}},\
  }\href@noop {} {\  (\bibinfo {year} {2019}{\natexlab{b}})},\ \Eprint
  {http://arxiv.org/abs/1910.09099} {arXiv:1910.09099 [astro-ph.CO]}
  \BibitemShut {NoStop}%
\bibitem [{\citenamefont {De~Luca}\ \emph {et~al.}(2019)\citenamefont
  {De~Luca}, \citenamefont {Franciolini}, \citenamefont {Kehagias},\ and\
  \citenamefont {Riotto}}]{DeLuca:2019ufz}%
  \BibitemOpen
  \bibfield  {author} {\bibinfo {author} {\bibfnamefont {V.}~\bibnamefont
  {De~Luca}}, \bibinfo {author} {\bibfnamefont {G.}~\bibnamefont
  {Franciolini}}, \bibinfo {author} {\bibfnamefont {A.}~\bibnamefont
  {Kehagias}}, \ and\ \bibinfo {author} {\bibfnamefont {A.}~\bibnamefont
  {Riotto}},\ }\href@noop {} {\  (\bibinfo {year} {2019})},\ \Eprint
  {http://arxiv.org/abs/1911.09689} {arXiv:1911.09689 [gr-qc]} \BibitemShut
  {NoStop}%
\bibitem [{\citenamefont {Tomikawa}\ and\ \citenamefont
  {Kobayashi}(2019)}]{Tomikawa:2019tvi}%
  \BibitemOpen
  \bibfield  {author} {\bibinfo {author} {\bibfnamefont {K.}~\bibnamefont
  {Tomikawa}}\ and\ \bibinfo {author} {\bibfnamefont {T.}~\bibnamefont
  {Kobayashi}},\ }\href@noop {} {\  (\bibinfo {year} {2019})},\ \Eprint
  {http://arxiv.org/abs/1910.01880} {arXiv:1910.01880 [gr-qc]} \BibitemShut
  {NoStop}%
\bibitem [{\citenamefont {Gong}(2019)}]{Gong:2019mui}%
  \BibitemOpen
  \bibfield  {author} {\bibinfo {author} {\bibfnamefont {J.-O.}\ \bibnamefont
  {Gong}},\ }\href@noop {} {\  (\bibinfo {year} {2019})},\ \Eprint
  {http://arxiv.org/abs/1909.12708} {arXiv:1909.12708 [gr-qc]} \BibitemShut
  {NoStop}%
\bibitem [{\citenamefont {Inomata}\ and\ \citenamefont
  {Terada}(2019)}]{Inomata:2019yww}%
  \BibitemOpen
  \bibfield  {author} {\bibinfo {author} {\bibfnamefont {K.}~\bibnamefont
  {Inomata}}\ and\ \bibinfo {author} {\bibfnamefont {T.}~\bibnamefont
  {Terada}},\ }\href@noop {} {\  (\bibinfo {year} {2019})},\ \Eprint
  {http://arxiv.org/abs/1912.00785} {arXiv:1912.00785 [gr-qc]} \BibitemShut
  {NoStop}%
\bibitem [{\citenamefont {Yuan}\ \emph
  {et~al.}(2019{\natexlab{c}})\citenamefont {Yuan}, \citenamefont {Chen},\ and\
  \citenamefont {Huang}}]{Yuan:2019fwv}%
  \BibitemOpen
  \bibfield  {author} {\bibinfo {author} {\bibfnamefont {C.}~\bibnamefont
  {Yuan}}, \bibinfo {author} {\bibfnamefont {Z.-C.}\ \bibnamefont {Chen}}, \
  and\ \bibinfo {author} {\bibfnamefont {Q.-G.}\ \bibnamefont {Huang}},\
  }\href@noop {} {\  (\bibinfo {year} {2019}{\natexlab{c}})},\ \Eprint
  {http://arxiv.org/abs/1912.00885} {arXiv:1912.00885 [astro-ph.CO]}
  \BibitemShut {NoStop}%
\bibitem [{\citenamefont {Watanabe}\ and\ \citenamefont
  {Komatsu}(2006)}]{Watanabe:2006qe}%
  \BibitemOpen
  \bibfield  {author} {\bibinfo {author} {\bibfnamefont {Y.}~\bibnamefont
  {Watanabe}}\ and\ \bibinfo {author} {\bibfnamefont {E.}~\bibnamefont
  {Komatsu}},\ }\href {\doibase 10.1103/PhysRevD.73.123515} {\bibfield
  {journal} {\bibinfo  {journal} {Phys. Rev.}\ }\textbf {\bibinfo {volume}
  {D73}},\ \bibinfo {pages} {123515} (\bibinfo {year} {2006})},\ \Eprint
  {http://arxiv.org/abs/astro-ph/0604176} {arXiv:astro-ph/0604176 [astro-ph]}
  \BibitemShut {NoStop}%
\bibitem [{\citenamefont {Cui}\ \emph {et~al.}(2018)\citenamefont {Cui},
  \citenamefont {Lewicki}, \citenamefont {Morrissey},\ and\ \citenamefont
  {Wells}}]{Cui:2017ufi}%
  \BibitemOpen
  \bibfield  {author} {\bibinfo {author} {\bibfnamefont {Y.}~\bibnamefont
  {Cui}}, \bibinfo {author} {\bibfnamefont {M.}~\bibnamefont {Lewicki}},
  \bibinfo {author} {\bibfnamefont {D.~E.}\ \bibnamefont {Morrissey}}, \ and\
  \bibinfo {author} {\bibfnamefont {J.~D.}\ \bibnamefont {Wells}},\ }\href
  {\doibase 10.1103/PhysRevD.97.123505} {\bibfield  {journal} {\bibinfo
  {journal} {Phys. Rev.}\ }\textbf {\bibinfo {volume} {D97}},\ \bibinfo {pages}
  {123505} (\bibinfo {year} {2018})},\ \Eprint
  {http://arxiv.org/abs/1711.03104} {arXiv:1711.03104 [hep-ph]} \BibitemShut
  {NoStop}%
\bibitem [{\citenamefont {Bettoni}\ \emph {et~al.}(2019)\citenamefont
  {Bettoni}, \citenamefont {Domènech},\ and\ \citenamefont
  {Rubio}}]{Bettoni:2018pbl}%
  \BibitemOpen
  \bibfield  {author} {\bibinfo {author} {\bibfnamefont {D.}~\bibnamefont
  {Bettoni}}, \bibinfo {author} {\bibfnamefont {G.}~\bibnamefont {Domènech}},
  \ and\ \bibinfo {author} {\bibfnamefont {J.}~\bibnamefont {Rubio}},\ }\href
  {\doibase 10.1088/1475-7516/2019/02/034} {\bibfield  {journal} {\bibinfo
  {journal} {JCAP}\ }\textbf {\bibinfo {volume} {1902}},\ \bibinfo {pages}
  {034} (\bibinfo {year} {2019})},\ \Eprint {http://arxiv.org/abs/1810.11117}
  {arXiv:1810.11117 [astro-ph.CO]} \BibitemShut {NoStop}%
\bibitem [{\citenamefont {Cui}\ \emph {et~al.}(2019)\citenamefont {Cui},
  \citenamefont {Lewicki}, \citenamefont {Morrissey},\ and\ \citenamefont
  {Wells}}]{Cui:2018rwi}%
  \BibitemOpen
  \bibfield  {author} {\bibinfo {author} {\bibfnamefont {Y.}~\bibnamefont
  {Cui}}, \bibinfo {author} {\bibfnamefont {M.}~\bibnamefont {Lewicki}},
  \bibinfo {author} {\bibfnamefont {D.~E.}\ \bibnamefont {Morrissey}}, \ and\
  \bibinfo {author} {\bibfnamefont {J.~D.}\ \bibnamefont {Wells}},\ }\href
  {\doibase 10.1007/JHEP01(2019)081} {\bibfield  {journal} {\bibinfo  {journal}
  {JHEP}\ }\textbf {\bibinfo {volume} {01}},\ \bibinfo {pages} {081} (\bibinfo
  {year} {2019})},\ \Eprint {http://arxiv.org/abs/1808.08968} {arXiv:1808.08968
  [hep-ph]} \BibitemShut {NoStop}%
\bibitem [{\citenamefont {Domènech}\ and\ \citenamefont
  {Sasaki}(2018)}]{Domenech:2017ems}%
  \BibitemOpen
  \bibfield  {author} {\bibinfo {author} {\bibfnamefont {G.}~\bibnamefont
  {Domènech}}\ and\ \bibinfo {author} {\bibfnamefont {M.}~\bibnamefont
  {Sasaki}},\ }\href {\doibase 10.1103/PhysRevD.97.023521} {\bibfield
  {journal} {\bibinfo  {journal} {Phys. Rev.}\ }\textbf {\bibinfo {volume}
  {D97}},\ \bibinfo {pages} {023521} (\bibinfo {year} {2018})},\ \Eprint
  {http://arxiv.org/abs/1709.09804} {arXiv:1709.09804 [gr-qc]} \BibitemShut
  {NoStop}%
\bibitem [{\citenamefont {Sasaki}\ \emph {et~al.}(2016)\citenamefont {Sasaki},
  \citenamefont {Suyama}, \citenamefont {Tanaka},\ and\ \citenamefont
  {Yokoyama}}]{Sasaki:2016jop}%
  \BibitemOpen
  \bibfield  {author} {\bibinfo {author} {\bibfnamefont {M.}~\bibnamefont
  {Sasaki}}, \bibinfo {author} {\bibfnamefont {T.}~\bibnamefont {Suyama}},
  \bibinfo {author} {\bibfnamefont {T.}~\bibnamefont {Tanaka}}, \ and\ \bibinfo
  {author} {\bibfnamefont {S.}~\bibnamefont {Yokoyama}},\ }\href {\doibase
  10.1103/PhysRevLett.121.059901, 10.1103/PhysRevLett.117.061101} {\bibfield
  {journal} {\bibinfo  {journal} {Phys. Rev. Lett.}\ }\textbf {\bibinfo
  {volume} {117}},\ \bibinfo {pages} {061101} (\bibinfo {year} {2016})},\
  \bibinfo {note} {[erratum: Phys. Rev. Lett.121,no.5,059901(2018)]},\ \Eprint
  {http://arxiv.org/abs/1603.08338} {arXiv:1603.08338 [astro-ph.CO]}
  \BibitemShut {NoStop}%
\bibitem [{\citenamefont {Bird}\ \emph {et~al.}(2016)\citenamefont {Bird},
  \citenamefont {Cholis}, \citenamefont {Muñoz}, \citenamefont {Ali-Haïmoud},
  \citenamefont {Kamionkowski}, \citenamefont {Kovetz}, \citenamefont
  {Raccanelli},\ and\ \citenamefont {Riess}}]{Bird:2016dcv}%
  \BibitemOpen
  \bibfield  {author} {\bibinfo {author} {\bibfnamefont {S.}~\bibnamefont
  {Bird}}, \bibinfo {author} {\bibfnamefont {I.}~\bibnamefont {Cholis}},
  \bibinfo {author} {\bibfnamefont {J.~B.}\ \bibnamefont {Muñoz}}, \bibinfo
  {author} {\bibfnamefont {Y.}~\bibnamefont {Ali-Haïmoud}}, \bibinfo {author}
  {\bibfnamefont {M.}~\bibnamefont {Kamionkowski}}, \bibinfo {author}
  {\bibfnamefont {E.~D.}\ \bibnamefont {Kovetz}}, \bibinfo {author}
  {\bibfnamefont {A.}~\bibnamefont {Raccanelli}}, \ and\ \bibinfo {author}
  {\bibfnamefont {A.~G.}\ \bibnamefont {Riess}},\ }\href {\doibase
  10.1103/PhysRevLett.116.201301} {\bibfield  {journal} {\bibinfo  {journal}
  {Phys. Rev. Lett.}\ }\textbf {\bibinfo {volume} {116}},\ \bibinfo {pages}
  {201301} (\bibinfo {year} {2016})},\ \Eprint
  {http://arxiv.org/abs/1603.00464} {arXiv:1603.00464 [astro-ph.CO]}
  \BibitemShut {NoStop}%
\bibitem [{\citenamefont {Carr}\ \emph {et~al.}(2016)\citenamefont {Carr},
  \citenamefont {Kuhnel},\ and\ \citenamefont {Sandstad}}]{Carr:2016drx}%
  \BibitemOpen
  \bibfield  {author} {\bibinfo {author} {\bibfnamefont {B.}~\bibnamefont
  {Carr}}, \bibinfo {author} {\bibfnamefont {F.}~\bibnamefont {Kuhnel}}, \ and\
  \bibinfo {author} {\bibfnamefont {M.}~\bibnamefont {Sandstad}},\ }\href
  {\doibase 10.1103/PhysRevD.94.083504} {\bibfield  {journal} {\bibinfo
  {journal} {Phys. Rev.}\ }\textbf {\bibinfo {volume} {D94}},\ \bibinfo {pages}
  {083504} (\bibinfo {year} {2016})},\ \Eprint
  {http://arxiv.org/abs/1607.06077} {arXiv:1607.06077 [astro-ph.CO]}
  \BibitemShut {NoStop}%
\bibitem [{\citenamefont {García-Bellido}(2017)}]{Garcia-Bellido:2017fdg}%
  \BibitemOpen
  \bibfield  {author} {\bibinfo {author} {\bibfnamefont {J.}~\bibnamefont
  {García-Bellido}},\ }\bibfield  {booktitle} {\emph {\bibinfo {booktitle}
  {{Proceedings, 11th International LISA Symposium: Zurich, Switzerland,
  September 5-9, 2016}}},\ }\href {\doibase 10.1088/1742-6596/840/1/012032}
  {\bibfield  {journal} {\bibinfo  {journal} {J. Phys. Conf. Ser.}\ }\textbf
  {\bibinfo {volume} {840}},\ \bibinfo {pages} {012032} (\bibinfo {year}
  {2017})},\ \Eprint {http://arxiv.org/abs/1702.08275} {arXiv:1702.08275
  [astro-ph.CO]} \BibitemShut {NoStop}%
\bibitem [{\citenamefont {{Sasaki}}\ \emph {et~al.}(2018)\citenamefont
  {{Sasaki}}, \citenamefont {Suyama}, \citenamefont {Tanaka},\ and\
  \citenamefont {Yokoyama}}]{Sasaki:2018dmp}%
  \BibitemOpen
  \bibfield  {author} {\bibinfo {author} {\bibfnamefont {M.}~\bibnamefont
  {{Sasaki}}}, \bibinfo {author} {\bibfnamefont {T.}~\bibnamefont {Suyama}},
  \bibinfo {author} {\bibfnamefont {T.}~\bibnamefont {Tanaka}}, \ and\ \bibinfo
  {author} {\bibfnamefont {S.}~\bibnamefont {Yokoyama}},\ }\href {\doibase
  10.1088/1361-6382/aaa7b4} {\bibfield  {journal} {\bibinfo  {journal} {Class.
  Quant. Grav.}\ }\textbf {\bibinfo {volume} {35}},\ \bibinfo {pages} {063001}
  (\bibinfo {year} {2018})},\ \Eprint {http://arxiv.org/abs/1801.05235}
  {arXiv:1801.05235 [astro-ph.CO]} \BibitemShut {NoStop}%
\bibitem [{\citenamefont {Nakama}\ \emph {et~al.}(2018)\citenamefont {Nakama},
  \citenamefont {Carr},\ and\ \citenamefont {Silk}}]{Nakama:2017xvq}%
  \BibitemOpen
  \bibfield  {author} {\bibinfo {author} {\bibfnamefont {T.}~\bibnamefont
  {Nakama}}, \bibinfo {author} {\bibfnamefont {B.}~\bibnamefont {Carr}}, \ and\
  \bibinfo {author} {\bibfnamefont {J.}~\bibnamefont {Silk}},\ }\href {\doibase
  10.1103/PhysRevD.97.043525} {\bibfield  {journal} {\bibinfo  {journal} {Phys.
  Rev.}\ }\textbf {\bibinfo {volume} {D97}},\ \bibinfo {pages} {043525}
  (\bibinfo {year} {2018})},\ \Eprint {http://arxiv.org/abs/1710.06945}
  {arXiv:1710.06945 [astro-ph.CO]} \BibitemShut {NoStop}%
\bibitem [{\citenamefont {Niikura}\ \emph {et~al.}(2019)\citenamefont {Niikura}
  \emph {et~al.}}]{Niikura:2017zjd}%
  \BibitemOpen
  \bibfield  {author} {\bibinfo {author} {\bibfnamefont {H.}~\bibnamefont
  {Niikura}} \emph {et~al.},\ }\href {\doibase 10.1038/s41550-019-0723-1}
  {\bibfield  {journal} {\bibinfo  {journal} {Nat. Astron.}\ }\textbf {\bibinfo
  {volume} {3}},\ \bibinfo {pages} {524} (\bibinfo {year} {2019})},\ \Eprint
  {http://arxiv.org/abs/1701.02151} {arXiv:1701.02151 [astro-ph.CO]}
  \BibitemShut {NoStop}%
\bibitem [{\citenamefont {Zumalacarregui}\ and\ \citenamefont
  {Seljak}(2018)}]{Zumalacarregui:2017qqd}%
  \BibitemOpen
  \bibfield  {author} {\bibinfo {author} {\bibfnamefont {M.}~\bibnamefont
  {Zumalacarregui}}\ and\ \bibinfo {author} {\bibfnamefont {U.}~\bibnamefont
  {Seljak}},\ }\href {\doibase 10.1103/PhysRevLett.121.141101} {\bibfield
  {journal} {\bibinfo  {journal} {Phys. Rev. Lett.}\ }\textbf {\bibinfo
  {volume} {121}},\ \bibinfo {pages} {141101} (\bibinfo {year} {2018})},\
  \Eprint {http://arxiv.org/abs/1712.02240} {arXiv:1712.02240 [astro-ph.CO]}
  \BibitemShut {NoStop}%
\bibitem [{\citenamefont {Gow}\ \emph {et~al.}(2019)\citenamefont {Gow},
  \citenamefont {Byrnes}, \citenamefont {Hall},\ and\ \citenamefont
  {Peacock}}]{Gow:2019pok}%
  \BibitemOpen
  \bibfield  {author} {\bibinfo {author} {\bibfnamefont {A.~D.}\ \bibnamefont
  {Gow}}, \bibinfo {author} {\bibfnamefont {C.~T.}\ \bibnamefont {Byrnes}},
  \bibinfo {author} {\bibfnamefont {A.}~\bibnamefont {Hall}}, \ and\ \bibinfo
  {author} {\bibfnamefont {J.~A.}\ \bibnamefont {Peacock}},\ }\href@noop {} {\
  (\bibinfo {year} {2019})},\ \Eprint {http://arxiv.org/abs/1911.12685}
  {arXiv:1911.12685 [astro-ph.CO]} \BibitemShut {NoStop}%
\bibitem [{\citenamefont {Bugaev}\ and\ \citenamefont
  {Klimai}(2010)}]{Bugaev:2009zh}%
  \BibitemOpen
  \bibfield  {author} {\bibinfo {author} {\bibfnamefont {E.}~\bibnamefont
  {Bugaev}}\ and\ \bibinfo {author} {\bibfnamefont {P.}~\bibnamefont
  {Klimai}},\ }\href {\doibase 10.1103/PhysRevD.81.023517} {\bibfield
  {journal} {\bibinfo  {journal} {Phys. Rev.}\ }\textbf {\bibinfo {volume}
  {D81}},\ \bibinfo {pages} {023517} (\bibinfo {year} {2010})},\ \Eprint
  {http://arxiv.org/abs/0908.0664} {arXiv:0908.0664 [astro-ph.CO]} \BibitemShut
  {NoStop}%
\bibitem [{\citenamefont {Saito}\ and\ \citenamefont
  {Yokoyama}(2010)}]{Saito:2009jt}%
  \BibitemOpen
  \bibfield  {author} {\bibinfo {author} {\bibfnamefont {R.}~\bibnamefont
  {Saito}}\ and\ \bibinfo {author} {\bibfnamefont {J.}~\bibnamefont
  {Yokoyama}},\ }\href {\doibase 10.1143/PTP.126.351, 10.1143/PTP.123.867}
  {\bibfield  {journal} {\bibinfo  {journal} {Prog. Theor. Phys.}\ }\textbf
  {\bibinfo {volume} {123}},\ \bibinfo {pages} {867} (\bibinfo {year}
  {2010})},\ \bibinfo {note} {[Erratum: Prog. Theor. Phys.126,351(2011)]},\
  \Eprint {http://arxiv.org/abs/0912.5317} {arXiv:0912.5317 [astro-ph.CO]}
  \BibitemShut {NoStop}%
\bibitem [{\citenamefont {Assadullahi}\ and\ \citenamefont
  {Wands}(2009)}]{Assadullahi:2009nf}%
  \BibitemOpen
  \bibfield  {author} {\bibinfo {author} {\bibfnamefont {H.}~\bibnamefont
  {Assadullahi}}\ and\ \bibinfo {author} {\bibfnamefont {D.}~\bibnamefont
  {Wands}},\ }\href {\doibase 10.1103/PhysRevD.79.083511} {\bibfield  {journal}
  {\bibinfo  {journal} {Phys. Rev.}\ }\textbf {\bibinfo {volume} {D79}},\
  \bibinfo {pages} {083511} (\bibinfo {year} {2009})},\ \Eprint
  {http://arxiv.org/abs/0901.0989} {arXiv:0901.0989 [astro-ph.CO]} \BibitemShut
  {NoStop}%
\bibitem [{\citenamefont {Wetterich}(1988)}]{Wetterich:1987fm}%
  \BibitemOpen
  \bibfield  {author} {\bibinfo {author} {\bibfnamefont {C.}~\bibnamefont
  {Wetterich}},\ }\href {\doibase 10.1016/0550-3213(88)90193-9} {\bibfield
  {journal} {\bibinfo  {journal} {Nucl. Phys.}\ }\textbf {\bibinfo {volume}
  {B302}},\ \bibinfo {pages} {668} (\bibinfo {year} {1988})},\ \Eprint
  {http://arxiv.org/abs/1711.03844} {arXiv:1711.03844 [hep-th]} \BibitemShut
  {NoStop}%
\bibitem [{\citenamefont {Spokoiny}(1993)}]{Spokoiny:1993kt}%
  \BibitemOpen
  \bibfield  {author} {\bibinfo {author} {\bibfnamefont {B.}~\bibnamefont
  {Spokoiny}},\ }\href {\doibase 10.1016/0370-2693(93)90155-B} {\bibfield
  {journal} {\bibinfo  {journal} {Phys. Lett.}\ }\textbf {\bibinfo {volume}
  {B315}},\ \bibinfo {pages} {40} (\bibinfo {year} {1993})},\ \Eprint
  {http://arxiv.org/abs/gr-qc/9306008} {arXiv:gr-qc/9306008 [gr-qc]}
  \BibitemShut {NoStop}%
\bibitem [{\citenamefont {Wetterich}(1995)}]{Wetterich:1994bg}%
  \BibitemOpen
  \bibfield  {author} {\bibinfo {author} {\bibfnamefont {C.}~\bibnamefont
  {Wetterich}},\ }\href@noop {} {\bibfield  {journal} {\bibinfo  {journal}
  {Astron. Astrophys.}\ }\textbf {\bibinfo {volume} {301}},\ \bibinfo {pages}
  {321} (\bibinfo {year} {1995})},\ \Eprint
  {http://arxiv.org/abs/hep-th/9408025} {arXiv:hep-th/9408025 [hep-th]}
  \BibitemShut {NoStop}%
\bibitem [{\citenamefont {Peebles}\ and\ \citenamefont
  {Vilenkin}(1999)}]{Peebles:1998qn}%
  \BibitemOpen
  \bibfield  {author} {\bibinfo {author} {\bibfnamefont {P.~J.~E.}\
  \bibnamefont {Peebles}}\ and\ \bibinfo {author} {\bibfnamefont
  {A.}~\bibnamefont {Vilenkin}},\ }\href {\doibase 10.1103/PhysRevD.59.063505}
  {\bibfield  {journal} {\bibinfo  {journal} {Phys. Rev.}\ }\textbf {\bibinfo
  {volume} {D59}},\ \bibinfo {pages} {063505} (\bibinfo {year} {1999})},\
  \Eprint {http://arxiv.org/abs/astro-ph/9810509} {arXiv:astro-ph/9810509
  [astro-ph]} \BibitemShut {NoStop}%
\bibitem [{\citenamefont {Brax}\ and\ \citenamefont
  {Martin}(2005)}]{Brax:2005uf}%
  \BibitemOpen
  \bibfield  {author} {\bibinfo {author} {\bibfnamefont {P.}~\bibnamefont
  {Brax}}\ and\ \bibinfo {author} {\bibfnamefont {J.}~\bibnamefont {Martin}},\
  }\href {\doibase 10.1103/PhysRevD.71.063530} {\bibfield  {journal} {\bibinfo
  {journal} {Phys. Rev.}\ }\textbf {\bibinfo {volume} {D71}},\ \bibinfo {pages}
  {063530} (\bibinfo {year} {2005})},\ \Eprint
  {http://arxiv.org/abs/astro-ph/0502069} {arXiv:astro-ph/0502069 [astro-ph]}
  \BibitemShut {NoStop}%
\bibitem [{\citenamefont {Wetterich}(2015)}]{Wetterich:2014gaa}%
  \BibitemOpen
  \bibfield  {author} {\bibinfo {author} {\bibfnamefont {C.}~\bibnamefont
  {Wetterich}},\ }\href {\doibase 10.1016/j.nuclphysb.2015.05.019} {\bibfield
  {journal} {\bibinfo  {journal} {Nucl. Phys.}\ }\textbf {\bibinfo {volume}
  {B897}},\ \bibinfo {pages} {111} (\bibinfo {year} {2015})},\ \Eprint
  {http://arxiv.org/abs/1408.0156} {arXiv:1408.0156 [hep-th]} \BibitemShut
  {NoStop}%
\bibitem [{\citenamefont {Wali~Hossain}\ \emph {et~al.}(2015)\citenamefont
  {Wali~Hossain}, \citenamefont {Myrzakulov}, \citenamefont {Sami},\ and\
  \citenamefont {Saridakis}}]{Hossain:2014zma}%
  \BibitemOpen
  \bibfield  {author} {\bibinfo {author} {\bibfnamefont {M.}~\bibnamefont
  {Wali~Hossain}}, \bibinfo {author} {\bibfnamefont {R.}~\bibnamefont
  {Myrzakulov}}, \bibinfo {author} {\bibfnamefont {M.}~\bibnamefont {Sami}}, \
  and\ \bibinfo {author} {\bibfnamefont {E.~N.}\ \bibnamefont {Saridakis}},\
  }\href {\doibase 10.1142/S0218271815300141} {\bibfield  {journal} {\bibinfo
  {journal} {Int. J. Mod. Phys.}\ }\textbf {\bibinfo {volume} {D24}},\ \bibinfo
  {pages} {1530014} (\bibinfo {year} {2015})},\ \Eprint
  {http://arxiv.org/abs/1410.6100} {arXiv:1410.6100 [gr-qc]} \BibitemShut
  {NoStop}%
\bibitem [{\citenamefont {Hossain}\ \emph {et~al.}(2014)\citenamefont
  {Hossain}, \citenamefont {Myrzakulov}, \citenamefont {Sami},\ and\
  \citenamefont {Saridakis}}]{Hossain:2014xha}%
  \BibitemOpen
  \bibfield  {author} {\bibinfo {author} {\bibfnamefont {M.~W.}\ \bibnamefont
  {Hossain}}, \bibinfo {author} {\bibfnamefont {R.}~\bibnamefont {Myrzakulov}},
  \bibinfo {author} {\bibfnamefont {M.}~\bibnamefont {Sami}}, \ and\ \bibinfo
  {author} {\bibfnamefont {E.~N.}\ \bibnamefont {Saridakis}},\ }\href {\doibase
  10.1103/PhysRevD.90.023512} {\bibfield  {journal} {\bibinfo  {journal} {Phys.
  Rev.}\ }\textbf {\bibinfo {volume} {D90}},\ \bibinfo {pages} {023512}
  (\bibinfo {year} {2014})},\ \Eprint {http://arxiv.org/abs/1402.6661}
  {arXiv:1402.6661 [gr-qc]} \BibitemShut {NoStop}%
\bibitem [{\citenamefont {Karananas}\ and\ \citenamefont
  {Rubio}(2016)}]{Karananas:2016kyt}%
  \BibitemOpen
  \bibfield  {author} {\bibinfo {author} {\bibfnamefont {G.~K.}\ \bibnamefont
  {Karananas}}\ and\ \bibinfo {author} {\bibfnamefont {J.}~\bibnamefont
  {Rubio}},\ }\href {\doibase 10.1016/j.physletb.2016.08.037} {\bibfield
  {journal} {\bibinfo  {journal} {Phys. Lett.}\ }\textbf {\bibinfo {volume}
  {B761}},\ \bibinfo {pages} {223} (\bibinfo {year} {2016})},\ \Eprint
  {http://arxiv.org/abs/1606.08848} {arXiv:1606.08848 [hep-ph]} \BibitemShut
  {NoStop}%
\bibitem [{\citenamefont {Agarwal}\ \emph {et~al.}(2017)\citenamefont
  {Agarwal}, \citenamefont {Myrzakulov}, \citenamefont {Sami},\ and\
  \citenamefont {Singh}}]{Agarwal:2017wxo}%
  \BibitemOpen
  \bibfield  {author} {\bibinfo {author} {\bibfnamefont {A.}~\bibnamefont
  {Agarwal}}, \bibinfo {author} {\bibfnamefont {R.}~\bibnamefont {Myrzakulov}},
  \bibinfo {author} {\bibfnamefont {M.}~\bibnamefont {Sami}}, \ and\ \bibinfo
  {author} {\bibfnamefont {N.~K.}\ \bibnamefont {Singh}},\ }\href {\doibase
  10.1016/j.physletb.2017.04.066} {\bibfield  {journal} {\bibinfo  {journal}
  {Phys. Lett.}\ }\textbf {\bibinfo {volume} {B770}},\ \bibinfo {pages} {200}
  (\bibinfo {year} {2017})},\ \Eprint {http://arxiv.org/abs/1708.00156}
  {arXiv:1708.00156 [gr-qc]} \BibitemShut {NoStop}%
\bibitem [{\citenamefont {Geng}\ \emph {et~al.}(2017)\citenamefont {Geng},
  \citenamefont {Lee}, \citenamefont {Sami}, \citenamefont {Saridakis},\ and\
  \citenamefont {Starobinsky}}]{Geng:2017mic}%
  \BibitemOpen
  \bibfield  {author} {\bibinfo {author} {\bibfnamefont {C.-Q.}\ \bibnamefont
  {Geng}}, \bibinfo {author} {\bibfnamefont {C.-C.}\ \bibnamefont {Lee}},
  \bibinfo {author} {\bibfnamefont {M.}~\bibnamefont {Sami}}, \bibinfo {author}
  {\bibfnamefont {E.~N.}\ \bibnamefont {Saridakis}}, \ and\ \bibinfo {author}
  {\bibfnamefont {A.~A.}\ \bibnamefont {Starobinsky}},\ }\href {\doibase
  10.1088/1475-7516/2017/06/011} {\bibfield  {journal} {\bibinfo  {journal}
  {JCAP}\ }\textbf {\bibinfo {volume} {1706}},\ \bibinfo {pages} {011}
  (\bibinfo {year} {2017})},\ \Eprint {http://arxiv.org/abs/1705.01329}
  {arXiv:1705.01329 [gr-qc]} \BibitemShut {NoStop}%
\bibitem [{\citenamefont {Dimopoulos}\ and\ \citenamefont
  {Owen}(2017)}]{Dimopoulos:2017zvq}%
  \BibitemOpen
  \bibfield  {author} {\bibinfo {author} {\bibfnamefont {K.}~\bibnamefont
  {Dimopoulos}}\ and\ \bibinfo {author} {\bibfnamefont {C.}~\bibnamefont
  {Owen}},\ }\href {\doibase 10.1088/1475-7516/2017/06/027} {\bibfield
  {journal} {\bibinfo  {journal} {JCAP}\ }\textbf {\bibinfo {volume} {1706}},\
  \bibinfo {pages} {027} (\bibinfo {year} {2017})},\ \Eprint
  {http://arxiv.org/abs/1703.00305} {arXiv:1703.00305 [gr-qc]} \BibitemShut
  {NoStop}%
\bibitem [{\citenamefont {Rubio}\ and\ \citenamefont
  {Wetterich}(2017)}]{Rubio:2017gty}%
  \BibitemOpen
  \bibfield  {author} {\bibinfo {author} {\bibfnamefont {J.}~\bibnamefont
  {Rubio}}\ and\ \bibinfo {author} {\bibfnamefont {C.}~\bibnamefont
  {Wetterich}},\ }\href {\doibase 10.1103/PhysRevD.96.063509} {\bibfield
  {journal} {\bibinfo  {journal} {Phys. Rev.}\ }\textbf {\bibinfo {volume}
  {D96}},\ \bibinfo {pages} {063509} (\bibinfo {year} {2017})},\ \Eprint
  {http://arxiv.org/abs/1705.00552} {arXiv:1705.00552 [gr-qc]} \BibitemShut
  {NoStop}%
\bibitem [{\citenamefont {Haro}\ \emph {et~al.}(2019)\citenamefont {Haro},
  \citenamefont {Yang},\ and\ \citenamefont {Pan}}]{Haro:2018zdb}%
  \BibitemOpen
  \bibfield  {author} {\bibinfo {author} {\bibfnamefont {J.}~\bibnamefont
  {Haro}}, \bibinfo {author} {\bibfnamefont {W.}~\bibnamefont {Yang}}, \ and\
  \bibinfo {author} {\bibfnamefont {S.}~\bibnamefont {Pan}},\ }\href {\doibase
  10.1088/1475-7516/2019/01/023} {\bibfield  {journal} {\bibinfo  {journal}
  {JCAP}\ }\textbf {\bibinfo {volume} {1901}},\ \bibinfo {pages} {023}
  (\bibinfo {year} {2019})},\ \Eprint {http://arxiv.org/abs/1811.07371}
  {arXiv:1811.07371 [gr-qc]} \BibitemShut {NoStop}%
\bibitem [{\citenamefont {Garcia-Bellido}\ \emph {et~al.}(2011)\citenamefont
  {Garcia-Bellido}, \citenamefont {Rubio}, \citenamefont {Shaposhnikov},\ and\
  \citenamefont {Zenhausern}}]{GarciaBellido:2011de}%
  \BibitemOpen
  \bibfield  {author} {\bibinfo {author} {\bibfnamefont {J.}~\bibnamefont
  {Garcia-Bellido}}, \bibinfo {author} {\bibfnamefont {J.}~\bibnamefont
  {Rubio}}, \bibinfo {author} {\bibfnamefont {M.}~\bibnamefont {Shaposhnikov}},
  \ and\ \bibinfo {author} {\bibfnamefont {D.}~\bibnamefont {Zenhausern}},\
  }\href {\doibase 10.1103/PhysRevD.84.123504} {\bibfield  {journal} {\bibinfo
  {journal} {Phys. Rev.}\ }\textbf {\bibinfo {volume} {D84}},\ \bibinfo {pages}
  {123504} (\bibinfo {year} {2011})},\ \Eprint {http://arxiv.org/abs/1107.2163}
  {arXiv:1107.2163 [hep-ph]} \BibitemShut {NoStop}%
\bibitem [{\citenamefont {Casas}\ \emph {et~al.}(2018)\citenamefont {Casas},
  \citenamefont {Pauly},\ and\ \citenamefont {Rubio}}]{Casas:2017wjh}%
  \BibitemOpen
  \bibfield  {author} {\bibinfo {author} {\bibfnamefont {S.}~\bibnamefont
  {Casas}}, \bibinfo {author} {\bibfnamefont {M.}~\bibnamefont {Pauly}}, \ and\
  \bibinfo {author} {\bibfnamefont {J.}~\bibnamefont {Rubio}},\ }\href
  {\doibase 10.1103/PhysRevD.97.043520} {\bibfield  {journal} {\bibinfo
  {journal} {Phys. Rev.}\ }\textbf {\bibinfo {volume} {D97}},\ \bibinfo {pages}
  {043520} (\bibinfo {year} {2018})},\ \Eprint
  {http://arxiv.org/abs/1712.04956} {arXiv:1712.04956 [astro-ph.CO]}
  \BibitemShut {NoStop}%
\bibitem [{\citenamefont {Byrnes}\ \emph {et~al.}(2018)\citenamefont {Byrnes},
  \citenamefont {Hindmarsh}, \citenamefont {Young},\ and\ \citenamefont
  {Hawkins}}]{Byrnes:2018clq}%
  \BibitemOpen
  \bibfield  {author} {\bibinfo {author} {\bibfnamefont {C.~T.}\ \bibnamefont
  {Byrnes}}, \bibinfo {author} {\bibfnamefont {M.}~\bibnamefont {Hindmarsh}},
  \bibinfo {author} {\bibfnamefont {S.}~\bibnamefont {Young}}, \ and\ \bibinfo
  {author} {\bibfnamefont {M.~R.~S.}\ \bibnamefont {Hawkins}},\ }\href
  {\doibase 10.1088/1475-7516/2018/08/041} {\bibfield  {journal} {\bibinfo
  {journal} {JCAP}\ }\textbf {\bibinfo {volume} {1808}},\ \bibinfo {pages}
  {041} (\bibinfo {year} {2018})},\ \Eprint {http://arxiv.org/abs/1801.06138}
  {arXiv:1801.06138 [astro-ph.CO]} \BibitemShut {NoStop}%
\bibitem [{\citenamefont {Hajkarim}\ and\ \citenamefont
  {Schaffner-Bielich}(2019)}]{Hajkarim:2019nbx}%
  \BibitemOpen
  \bibfield  {author} {\bibinfo {author} {\bibfnamefont {F.}~\bibnamefont
  {Hajkarim}}\ and\ \bibinfo {author} {\bibfnamefont {J.}~\bibnamefont
  {Schaffner-Bielich}},\ }\href@noop {} {\  (\bibinfo {year} {2019})},\ \Eprint
  {http://arxiv.org/abs/1910.12357} {arXiv:1910.12357 [hep-ph]} \BibitemShut
  {NoStop}%
\bibitem [{\citenamefont {Arroja}\ \emph {et~al.}(2009)\citenamefont {Arroja},
  \citenamefont {Assadullahi}, \citenamefont {Koyama},\ and\ \citenamefont
  {Wands}}]{Arroja:2009sh}%
  \BibitemOpen
  \bibfield  {author} {\bibinfo {author} {\bibfnamefont {F.}~\bibnamefont
  {Arroja}}, \bibinfo {author} {\bibfnamefont {H.}~\bibnamefont {Assadullahi}},
  \bibinfo {author} {\bibfnamefont {K.}~\bibnamefont {Koyama}}, \ and\ \bibinfo
  {author} {\bibfnamefont {D.}~\bibnamefont {Wands}},\ }\href {\doibase
  10.1103/PhysRevD.80.123526} {\bibfield  {journal} {\bibinfo  {journal} {Phys.
  Rev.}\ }\textbf {\bibinfo {volume} {D80}},\ \bibinfo {pages} {123526}
  (\bibinfo {year} {2009})},\ \Eprint {http://arxiv.org/abs/0907.3618}
  {arXiv:0907.3618 [astro-ph.CO]} \BibitemShut {NoStop}%
\bibitem [{\citenamefont {Mangilli}\ \emph {et~al.}(2008)\citenamefont
  {Mangilli}, \citenamefont {Bartolo}, \citenamefont {Matarrese},\ and\
  \citenamefont {Riotto}}]{Mangilli:2008bw}%
  \BibitemOpen
  \bibfield  {author} {\bibinfo {author} {\bibfnamefont {A.}~\bibnamefont
  {Mangilli}}, \bibinfo {author} {\bibfnamefont {N.}~\bibnamefont {Bartolo}},
  \bibinfo {author} {\bibfnamefont {S.}~\bibnamefont {Matarrese}}, \ and\
  \bibinfo {author} {\bibfnamefont {A.}~\bibnamefont {Riotto}},\ }\href
  {\doibase 10.1103/PhysRevD.78.083517} {\bibfield  {journal} {\bibinfo
  {journal} {Phys. Rev.}\ }\textbf {\bibinfo {volume} {D78}},\ \bibinfo {pages}
  {083517} (\bibinfo {year} {2008})},\ \Eprint {http://arxiv.org/abs/0805.3234}
  {arXiv:0805.3234 [astro-ph]} \BibitemShut {NoStop}%
\bibitem [{\citenamefont {Maggiore}(2007)}]{Maggiore:1900zz}%
  \BibitemOpen
  \bibfield  {author} {\bibinfo {author} {\bibfnamefont {M.}~\bibnamefont
  {Maggiore}},\ }\href {http://www.oup.com/uk/catalogue/?ci=9780198570745}
  {\emph {\bibinfo {title} {{Gravitational Waves. Vol. 1: Theory and
  Experiments}}}},\ Oxford Master Series in Physics\ (\bibinfo  {publisher}
  {Oxford University Press},\ \bibinfo {year} {2007})\BibitemShut {NoStop}%
\bibitem [{\citenamefont {Garcia-Bellido}\ \emph {et~al.}(2017)\citenamefont
  {Garcia-Bellido}, \citenamefont {Peloso},\ and\ \citenamefont
  {Unal}}]{Garcia-Bellido:2017aan}%
  \BibitemOpen
  \bibfield  {author} {\bibinfo {author} {\bibfnamefont {J.}~\bibnamefont
  {Garcia-Bellido}}, \bibinfo {author} {\bibfnamefont {M.}~\bibnamefont
  {Peloso}}, \ and\ \bibinfo {author} {\bibfnamefont {C.}~\bibnamefont
  {Unal}},\ }\href {\doibase 10.1088/1475-7516/2017/09/013} {\bibfield
  {journal} {\bibinfo  {journal} {JCAP}\ }\textbf {\bibinfo {volume} {1709}},\
  \bibinfo {pages} {013} (\bibinfo {year} {2017})},\ \Eprint
  {http://arxiv.org/abs/1707.02441} {arXiv:1707.02441 [astro-ph.CO]}
  \BibitemShut {NoStop}%
\bibitem [{\citenamefont {Unal}(2019)}]{Unal:2018yaa}%
  \BibitemOpen
  \bibfield  {author} {\bibinfo {author} {\bibfnamefont {C.}~\bibnamefont
  {Unal}},\ }\href {\doibase 10.1103/PhysRevD.99.041301} {\bibfield  {journal}
  {\bibinfo  {journal} {Phys. Rev.}\ }\textbf {\bibinfo {volume} {D99}},\
  \bibinfo {pages} {041301} (\bibinfo {year} {2019})},\ \Eprint
  {http://arxiv.org/abs/1811.09151} {arXiv:1811.09151 [astro-ph.CO]}
  \BibitemShut {NoStop}%
\bibitem [{\citenamefont {Bamba}\ \emph {et~al.}(2018)\citenamefont {Bamba},
  \citenamefont {Nojiri},\ and\ \citenamefont {Odintsov}}]{Bamba:2018cup}%
  \BibitemOpen
  \bibfield  {author} {\bibinfo {author} {\bibfnamefont {K.}~\bibnamefont
  {Bamba}}, \bibinfo {author} {\bibfnamefont {S.}~\bibnamefont {Nojiri}}, \
  and\ \bibinfo {author} {\bibfnamefont {S.~D.}\ \bibnamefont {Odintsov}},\
  }\href {\doibase 10.1103/PhysRevD.98.024002} {\bibfield  {journal} {\bibinfo
  {journal} {Phys. Rev.}\ }\textbf {\bibinfo {volume} {D98}},\ \bibinfo {pages}
  {024002} (\bibinfo {year} {2018})},\ \Eprint
  {http://arxiv.org/abs/1804.02275} {arXiv:1804.02275 [gr-qc]} \BibitemShut
  {NoStop}%
\bibitem [{\citenamefont {Gervois}\ and\ \citenamefont
  {Navelet}(1985)}]{threebesselI}%
  \BibitemOpen
  \bibfield  {author} {\bibinfo {author} {\bibfnamefont {A.}~\bibnamefont
  {Gervois}}\ and\ \bibinfo {author} {\bibfnamefont {H.}~\bibnamefont
  {Navelet}},\ }\href {\doibase 10.1063/1.526600} {\bibfield  {journal}
  {\bibinfo  {journal} {Journal of Mathematical Physics}\ }\textbf {\bibinfo
  {volume} {26}},\ \bibinfo {pages} {633} (\bibinfo {year} {1985})},\ \Eprint
  {http://arxiv.org/abs/https://doi.org/10.1063/1.526600}
  {https://doi.org/10.1063/1.526600} \BibitemShut {NoStop}%
\bibitem [{\citenamefont {Bailey}(1936)}]{originalbessel}%
  \BibitemOpen
  \bibfield  {author} {\bibinfo {author} {\bibfnamefont {W.~N.}\ \bibnamefont
  {Bailey}},\ }\href {\doibase 10.1112/plms/s2-40.1.37} {\bibfield  {journal}
  {\bibinfo  {journal} {Proceedings of the London Mathematical Society}\
  }\textbf {\bibinfo {volume} {s2-40}},\ \bibinfo {pages} {37} (\bibinfo {year}
  {1936})},\ \Eprint
  {http://arxiv.org/abs/https://londmathsoc.onlinelibrary.wiley.com/doi/pdf/10.1112/plms/s2-40.1.37}
  {https://londmathsoc.onlinelibrary.wiley.com/doi/pdf/10.1112/plms/s2-40.1.37}
  \BibitemShut {NoStop}%
\bibitem [{{\relax DLMF}()}]{NIST:DLMF}%
  \BibitemOpen
  {\relax DLMF},\ \href {http://dlmf.nist.gov/} {\enquote {\bibinfo {title}
  {{\it NIST Digital Library of Mathematical Functions}},}\ }\bibinfo
  {howpublished} {http://dlmf.nist.gov/, Release 1.0.24 of 2019-09-15},\
  \bibinfo {note} {f.~W.~J. Olver, A.~B. {Olde Daalhuis}, D.~W. Lozier, B.~I.
  Schneider, R.~F. Boisvert, C.~W. Clark, B.~R. Miller, B.~V. Saunders, H.~S.
  Cohl, and M.~A. McClain, eds.}\BibitemShut {Stop}%
\bibitem [{\citenamefont {Cai}\ \emph {et~al.}(2019{\natexlab{b}})\citenamefont
  {Cai}, \citenamefont {Pi},\ and\ \citenamefont {Sasaki}}]{Cai:2019cdl}%
  \BibitemOpen
  \bibfield  {author} {\bibinfo {author} {\bibfnamefont {R.-G.}\ \bibnamefont
  {Cai}}, \bibinfo {author} {\bibfnamefont {S.}~\bibnamefont {Pi}}, \ and\
  \bibinfo {author} {\bibfnamefont {M.}~\bibnamefont {Sasaki}},\ }\href@noop {}
  {\  (\bibinfo {year} {2019}{\natexlab{b}})},\ \Eprint
  {http://arxiv.org/abs/1909.13728} {arXiv:1909.13728 [astro-ph.CO]}
  \BibitemShut {NoStop}%
\bibitem [{\citenamefont {Carr}\ \emph {et~al.}(2010)\citenamefont {Carr},
  \citenamefont {Kohri}, \citenamefont {Sendouda},\ and\ \citenamefont
  {Yokoyama}}]{Carr:2009jm}%
  \BibitemOpen
  \bibfield  {author} {\bibinfo {author} {\bibfnamefont {B.~J.}\ \bibnamefont
  {Carr}}, \bibinfo {author} {\bibfnamefont {K.}~\bibnamefont {Kohri}},
  \bibinfo {author} {\bibfnamefont {Y.}~\bibnamefont {Sendouda}}, \ and\
  \bibinfo {author} {\bibfnamefont {J.}~\bibnamefont {Yokoyama}},\ }\href
  {\doibase 10.1103/PhysRevD.81.104019} {\bibfield  {journal} {\bibinfo
  {journal} {Phys. Rev.}\ }\textbf {\bibinfo {volume} {D81}},\ \bibinfo {pages}
  {104019} (\bibinfo {year} {2010})},\ \Eprint {http://arxiv.org/abs/0912.5297}
  {arXiv:0912.5297 [astro-ph.CO]} \BibitemShut {NoStop}%
\bibitem [{\citenamefont {Harada}\ \emph {et~al.}(2013)\citenamefont {Harada},
  \citenamefont {Yoo},\ and\ \citenamefont {Kohri}}]{Harada:2013epa}%
  \BibitemOpen
  \bibfield  {author} {\bibinfo {author} {\bibfnamefont {T.}~\bibnamefont
  {Harada}}, \bibinfo {author} {\bibfnamefont {C.-M.}\ \bibnamefont {Yoo}}, \
  and\ \bibinfo {author} {\bibfnamefont {K.}~\bibnamefont {Kohri}},\ }\href
  {\doibase 10.1103/PhysRevD.88.084051, 10.1103/PhysRevD.89.029903} {\bibfield
  {journal} {\bibinfo  {journal} {Phys. Rev.}\ }\textbf {\bibinfo {volume}
  {D88}},\ \bibinfo {pages} {084051} (\bibinfo {year} {2013})},\ \bibinfo
  {note} {[Erratum: Phys. Rev.D89,no.2,029903(2014)]},\ \Eprint
  {http://arxiv.org/abs/1309.4201} {arXiv:1309.4201 [astro-ph.CO]} \BibitemShut
  {NoStop}%
\bibitem [{\citenamefont {Escriv\`a}\ \emph {et~al.}(2019)\citenamefont
  {Escriv\`a}, \citenamefont {Germani},\ and\ \citenamefont
  {Sheth}}]{Escriva:2019phb}%
  \BibitemOpen
  \bibfield  {author} {\bibinfo {author} {\bibfnamefont {A.}~\bibnamefont
  {Escriv\`a}}, \bibinfo {author} {\bibfnamefont {C.}~\bibnamefont {Germani}},
  \ and\ \bibinfo {author} {\bibfnamefont {R.~K.}\ \bibnamefont {Sheth}},\
  }\href@noop {} {\  (\bibinfo {year} {2019})},\ \Eprint
  {http://arxiv.org/abs/1907.13311} {arXiv:1907.13311 [gr-qc]} \BibitemShut
  {NoStop}%
\bibitem [{\citenamefont {Musco}\ and\ \citenamefont
  {Miller}(2013)}]{Musco:2012au}%
  \BibitemOpen
  \bibfield  {author} {\bibinfo {author} {\bibfnamefont {I.}~\bibnamefont
  {Musco}}\ and\ \bibinfo {author} {\bibfnamefont {J.~C.}\ \bibnamefont
  {Miller}},\ }\href {\doibase 10.1088/0264-9381/30/14/145009} {\bibfield
  {journal} {\bibinfo  {journal} {Class. Quant. Grav.}\ }\textbf {\bibinfo
  {volume} {30}},\ \bibinfo {pages} {145009} (\bibinfo {year} {2013})},\
  \Eprint {http://arxiv.org/abs/1201.2379} {arXiv:1201.2379 [gr-qc]}
  \BibitemShut {NoStop}%
\bibitem [{\citenamefont {Escrivà}(2019)}]{Escriva:2019nsa}%
  \BibitemOpen
  \bibfield  {author} {\bibinfo {author} {\bibfnamefont {A.}~\bibnamefont
  {Escrivà}},\ }\href@noop {} {\  (\bibinfo {year} {2019})},\ \Eprint
  {http://arxiv.org/abs/1907.13065} {arXiv:1907.13065 [gr-qc]} \BibitemShut
  {NoStop}%
\bibitem [{\citenamefont {Germani}\ and\ \citenamefont
  {Musco}(2019)}]{Germani:2018jgr}%
  \BibitemOpen
  \bibfield  {author} {\bibinfo {author} {\bibfnamefont {C.}~\bibnamefont
  {Germani}}\ and\ \bibinfo {author} {\bibfnamefont {I.}~\bibnamefont
  {Musco}},\ }\href {\doibase 10.1103/PhysRevLett.122.141302} {\bibfield
  {journal} {\bibinfo  {journal} {Phys. Rev. Lett.}\ }\textbf {\bibinfo
  {volume} {122}},\ \bibinfo {pages} {141302} (\bibinfo {year} {2019})},\
  \Eprint {http://arxiv.org/abs/1805.04087} {arXiv:1805.04087 [astro-ph.CO]}
  \BibitemShut {NoStop}%
\bibitem [{\citenamefont {Atal}\ and\ \citenamefont
  {Germani}(2019)}]{Atal:2018neu}%
  \BibitemOpen
  \bibfield  {author} {\bibinfo {author} {\bibfnamefont {V.}~\bibnamefont
  {Atal}}\ and\ \bibinfo {author} {\bibfnamefont {C.}~\bibnamefont {Germani}},\
  }\href {\doibase 10.1016/j.dark.2019.100275} {\bibfield  {journal} {\bibinfo
  {journal} {Phys. Dark Univ.}\ }\textbf {\bibinfo {volume} {24}},\ \bibinfo
  {pages} {100275} (\bibinfo {year} {2019})},\ \Eprint
  {http://arxiv.org/abs/1811.07857} {arXiv:1811.07857 [astro-ph.CO]}
  \BibitemShut {NoStop}%
\bibitem [{\citenamefont {Young}\ \emph {et~al.}(2019)\citenamefont {Young},
  \citenamefont {Musco},\ and\ \citenamefont {Byrnes}}]{Young:2019yug}%
  \BibitemOpen
  \bibfield  {author} {\bibinfo {author} {\bibfnamefont {S.}~\bibnamefont
  {Young}}, \bibinfo {author} {\bibfnamefont {I.}~\bibnamefont {Musco}}, \ and\
  \bibinfo {author} {\bibfnamefont {C.~T.}\ \bibnamefont {Byrnes}},\ }\href
  {\doibase 10.1088/1475-7516/2019/11/012} {\bibfield  {journal} {\bibinfo
  {journal} {JCAP}\ }\textbf {\bibinfo {volume} {1911}},\ \bibinfo {pages}
  {012} (\bibinfo {year} {2019})},\ \Eprint {http://arxiv.org/abs/1904.00984}
  {arXiv:1904.00984 [astro-ph.CO]} \BibitemShut {NoStop}%
\bibitem [{\citenamefont {Atal}\ \emph {et~al.}(2019)\citenamefont {Atal},
  \citenamefont {Garriga},\ and\ \citenamefont
  {Marcos-Caballero}}]{Atal:2019cdz}%
  \BibitemOpen
  \bibfield  {author} {\bibinfo {author} {\bibfnamefont {V.}~\bibnamefont
  {Atal}}, \bibinfo {author} {\bibfnamefont {J.}~\bibnamefont {Garriga}}, \
  and\ \bibinfo {author} {\bibfnamefont {A.}~\bibnamefont {Marcos-Caballero}},\
  }\href@noop {} {\  (\bibinfo {year} {2019})},\ \Eprint
  {http://arxiv.org/abs/1905.13202} {arXiv:1905.13202 [astro-ph.CO]}
  \BibitemShut {NoStop}%
\bibitem [{\citenamefont {Thrane}\ and\ \citenamefont
  {Romano}(2013)}]{Thrane:2013oya}%
  \BibitemOpen
  \bibfield  {author} {\bibinfo {author} {\bibfnamefont {E.}~\bibnamefont
  {Thrane}}\ and\ \bibinfo {author} {\bibfnamefont {J.~D.}\ \bibnamefont
  {Romano}},\ }\href {\doibase 10.1103/PhysRevD.88.124032} {\bibfield
  {journal} {\bibinfo  {journal} {Phys. Rev.}\ }\textbf {\bibinfo {volume}
  {D88}},\ \bibinfo {pages} {124032} (\bibinfo {year} {2013})},\ \Eprint
  {http://arxiv.org/abs/1310.5300} {arXiv:1310.5300 [astro-ph.IM]} \BibitemShut
  {NoStop}%
\bibitem [{\citenamefont {Moore}\ \emph {et~al.}(2015)\citenamefont {Moore},
  \citenamefont {Cole},\ and\ \citenamefont {Berry}}]{Moore:2014lga}%
  \BibitemOpen
  \bibfield  {author} {\bibinfo {author} {\bibfnamefont {C.~J.}\ \bibnamefont
  {Moore}}, \bibinfo {author} {\bibfnamefont {R.~H.}\ \bibnamefont {Cole}}, \
  and\ \bibinfo {author} {\bibfnamefont {C.~P.~L.}\ \bibnamefont {Berry}},\
  }\href {\doibase 10.1088/0264-9381/32/1/015014} {\bibfield  {journal}
  {\bibinfo  {journal} {Class. Quant. Grav.}\ }\textbf {\bibinfo {volume}
  {32}},\ \bibinfo {pages} {015014} (\bibinfo {year} {2015})},\ \Eprint
  {http://arxiv.org/abs/1408.0740} {arXiv:1408.0740 [gr-qc]} \BibitemShut
  {NoStop}%
\bibitem [{\citenamefont {Young}(2019)}]{Young:2019osy}%
  \BibitemOpen
  \bibfield  {author} {\bibinfo {author} {\bibfnamefont {S.}~\bibnamefont
  {Young}},\ }\href@noop {} {\  (\bibinfo {year} {2019})},\ \Eprint
  {http://arxiv.org/abs/1905.01230} {arXiv:1905.01230 [astro-ph.CO]}
  \BibitemShut {NoStop}%
\end{thebibliography}%

\end{document}